\pdfoutput=1

\documentclass[12pt,a4paper]{article}

\usepackage{jheplike,ifpdf,tocloft,rotating}

\hypersetup{pdftitle={},pdfcreator={},linkcolor=[rgb]{0.15,0.35,0.75},colorlinks=true,citecolor=[rgb]{0.675,0,0.2},urlcolor=[rgb]{0.15,0.35,0.65}}
\graphicspath{{figures/}}

\let\olditemize\itemize\renewcommand{\itemize}{\vspace{-2pt}\olditemize\setlength{\itemsep}{1pt}\setlength{\parskip}{0pt}\setlength{\parsep}{-0pt}}
\let\oldenumerate\enumerate\renewcommand{\enumerate}{\vspace{-4pt}\oldenumerate\setlength{\itemsep}{1pt}\setlength{\parskip}{0pt}\setlength{\parsep}{0pt}}

\newcommand{\eq}[1]{\vspace{-0.5pt}\begin{equation}#1\vspace{-0.5pt}\end{equation}}

\newcommand{\fwbox}[2]{\text{\makebox[#1][c]{$\hspace{-150pt}\displaystyle#2\hspace{-150pt}$}}}
\newcommand{\fwboxL}[2]{\text{\makebox[#1][l]{$#2$}}}
\newcommand{\fwboxR}[2]{\text{\makebox[#1][r]{$#2$}}}
\newcommand{\fig}[3]{\raisebox{#1}{\includegraphics[scale=#2]{#3}}}
\newcommand{\mi}{\raisebox{0.75pt}{\scalebox{0.75}{$\,-\,$}}}
\newcommand{\pl}{\raisebox{0.75pt}{\scalebox{0.75}{$\,+\,$}}}

\newcommand{\x}[2]{(#1,#2)}

\newcommand{\bigger}[1]{\raisebox{-2.25pt}{\scalebox{1.75}{$#1$}}}

\DeclareMathOperator*{\Res}{\mathrm{Res}}
\definecolor{mhvBlue}{rgb}{0.3,0.2,0.75}
\definecolor{fRed}{rgb}{0.48,0.02824,0.18824}
\definecolor{cut2}{rgb}{0.18824,0.18824,0.48}
\definecolor{cut1}{rgb}{0.48,0.02824,0.18824}
\definecolor{hblue}{rgb}{0,0,0.575}
\definecolor{hred}{rgb}{0.475,0.0,0.15}
\definecolor{dred}{rgb}{0.575,0.4,0.45}

\title{~\\[15pt]{\LARGE \mbox{Prescriptive Unitarity}}\\[-24pt]}
\author[1,4]{\vspace{-10pt}Jacob~L.~Bourjaily}\affiliation[1]{Niels Bohr International Academy and Discovery Center, University of Copenhagen\\The Niels Bohr Institute, Blegdamsvej 17, DK-2100, Copenhagen \O, Denmark}\emailAdd{bourjaily@nbi.ku.dk}
\author[2,4]{\!\!\!,\,Enrico~Herrmann}\affiliation[2]{Walter Burke Institute for Theoretical Physics,\\ California Institute of Technology, Pasadena, CA 91125, USA}\emailAdd{eherrmann@caltech.edu}
\author[3,4]{\!\!\!,\,Jaroslav~Trnka}\affiliation[3]{Center for Quantum Mathematics and Physics (QMAP),\\Department of Physics, University of California, Davis, CA 95616, USA}\emailAdd{trnka@ucdavis.edu}
\affiliation[4]{Kavli Institute for Theoretical Physics, \\University of California, Santa Barbara, CA 93106, USA}
\abstract{
We introduce a {\it prescriptive} approach to generalized unitarity, resulting in a strictly-diagonal basis of loop integrands with coefficients given by specifically-tailored residues in field theory. We illustrate the power of this strategy in the case of planar, maximally supersymmetric Yang-Mills theory (SYM), where we construct closed-form representations of all ($n$-point N$^k$MHV) scattering amplitudes through three loops. The prescriptive approach contrasts with the ordinary description of unitarity-based methods by avoiding any need for linear algebra to determine integrand coefficients. We describe this approach in general terms as it should have applications to many quantum field theories, including those without planarity, supersymmetry, or massless spectra defined in any number of dimensions.
}
\preprint{CALT-TH-2017-19}

\begin{document}\maketitle
\vspace{-0pt}\section{Introduction and Overview}\label{sec:introduction}\vspace{-8pt}

There has been tremendous progress in recent years in the calculation and understanding of perturbative scattering amplitudes in quantum field theory. The scope of these insights and the powerful new computational tools that have resulted include many unexpected connections to modern developments in mathematics, {\it e.g.}\ \cite{Lusztig,Postnikov:2006kva,Postnikov2009matching,Williams:2003a,Goncharov:2011hp,KLS}. Most of these discoveries have been fueled by direct computation---pushing the limits of our theoretical reach (often for toy models) to uncover unanticipated, simplifying structures in the formulae that result, and using these insights to build more powerful tools. The lessons learned through such investigations include the (BCFW) on-shell recursion relations at tree- and loop-level, \cite{BCF, BCFW} and \cite{ArkaniHamed:2010kv}; the discovery of a hidden dual conformal invariance \cite{Drummond:2006rz,Alday:2007hr,Drummond:2008vq} as well as the duality to Wilson loops and correlation functions \cite{Drummond:2007aua,Brandhuber:2007yx,Drummond:2007au,Mason:2010yk,CaronHuot:2010ek,Alday:2010zy,Eden:2010zz,Eden:2010ce}; the connection to Grassmannian geometry \cite{ArkaniHamed:2012nw,ArkaniHamed:2009dn,ArkaniHamed:2009vw,Mason:2009qx,ArkaniHamed:2009dg,ArkaniHamed:2009sx,Huang:2013owa} and the amplituhedron \cite{ArkaniHamed:2010gg,Arkani-Hamed:2013jha,Arkani-Hamed:2013kca,Franco:2014csa,Lam:2014jda,Bai:2014cna,Bai:2015qoa,Ferro:2015grk,Ferro:2016zmx,Galloni:2016iuj,Karp:2016uax,Arkani-Hamed:2017tmz}; various bootstrap methods \cite{Dixon:2011pw,Dixon:2011nj,Dixon:2013eka,Dixon:2014iba,Dixon:2015iva,Caron-Huot:2016owq,Drummond:2014ffa,Dixon:2016nkn,Goncharov:2010jf,Golden:2013xva,Parker:2015cia,Duhr:2011zq}; the twistor string \cite{Witten:2003nn,Roiban:2004vt} and its generalization to the scattering equation formalism \cite{Cachazo:2013hca,Cachazo:2013iea,Cachazo:2013gna,Cachazo:2014xea,Dolan:2014ega,Bjerrum-Bohr:2014qwa,Baadsgaard:2015hia,Baadsgaard:2015ifa,Mason:2013sva,Geyer:2015bja,Geyer:2014fka,Cachazo:2012da,Geyer:2015jch,Casali:2015vta,Geyer:2016wjx}; to $Q$-cuts \cite{Baadsgaard:2015twa,Huang:2015cwh,An:2016awq}, and so on. For a broad overview of some of these developments, see {\it e.g.}\ \cite{Bern:1996je,Cachazo:2005ga,Beisert:2010jr,Elvang:2013cua,Dixon:2013uaa,Henn:2014yza,Duhr:2014woa,Weinzierl:2016bus}.

This progress has been fueled by very concrete computational targets often guided by specific physical questions. Beyond improving our predictive reach for {\it e.g.}\ collider physics applications, such computations are often critical to theoretical investigations. These include the ultraviolet properties of quantum gravity \cite{Bern:1998ug,Bern:2007xj,Bern:2012uf,Bern:2015xsa} and the (mathematical) structures behind the functions and numbers that result from perturbation theory \cite{Broadhurst:1996kc,Kreimer:1997dp,Bloch:2005bh,Brown:2009ta,Brown:2010bw,Bogner:2014mha,Brown:2015fyf,Panzer:2015ida,Marcolli:2009zy,Aluffi:2008sy,Brown:2009rc}. More generally, these efforts are often motivated by the enormous discrepancy between the difficulty of computations in field theory and the profound simplicities of the predictions that ultimately result. Some of these simplicities, such as finiteness and the logarithmic behavior of loop integrands are known to be in tension with the ways we normally represent amplitudes (see {\it e.g.}\ \cite{Bern:2007xj,Bern:2012uf,Arkani-Hamed:2014via,Bern:2014kca,Bern:2015ple,Bourjaily:2011hi,Bourjaily:2015bpz,Bourjaily:2016evz}). Exposing such tension through direct computation can be very illuminating, often leading to new insights into what we hope will contribute to a better understanding of the foundations of quantum field theory. 

Of all the methods used to push the limits of our computational reach into perturbation theory, the most generally applicable is also perhaps the most universal: {\it generalized unitarity} (see {\it e.g.}\ \cite{Bern:1994zx,Bern:1994cg,Britto:2004nc,Anastasiou:2006jv,Bern:2007ct,Cachazo:2008vp,Berger:2008sj,Abreu:2017xsl}). The basic idea is very simple. Because loop {\it integrands} are rational functions, they should be determinable by their residues. And because the space of loop integrands---viewed as rational functions---is always finite-dimensional, we can expand any amplitude into a complete basis of such functions:
\vspace{-6.5pt}\eq{\mathcal{A}_n^{L}=\sum_{k} c_k\,\mathcal{I}_k\,.\label{generalized_unitarity_intro}\vspace{-5pt}}
(The size of this basis depends on the spacetime dimension and the power counting of the quantum field theory in question.) Given any complete basis $\{\mathcal{I}_k\}$, the coefficients $c_k$ of any loop amplitude $\mathcal{A}_n^{L}$ in (\ref{generalized_unitarity_intro}) can then be determined by linear algebra via the criterion that residues match field theory. 

This approach is quite general: it can be used to find a representation of {\it any} perturbative amplitude in {\it any} quantum field theory expanded into a canonical basis of fixed integrals.\footnote{There {are} subtleties for amplitudes that are not fully `cut constructible'; but these can always be addressed ({\it e.g.}\ via dimensional regularization, see {\it e.g.}\ \cite{Anastasiou:2006jv}), and will not concern us here.} For recent applications of the unitarity method as well as integrand based reduction algorithms, see {\it e.g.}\ \cite{Passarino:1978jh,Ossola:2006us,Mastrolia:2010nb,Badger:2012dv,Mastrolia:2012an,Badger:2013sta,Mastrolia:2016dhn,Ita:2015tya}. Among the most important advantages of this approach (relative to the Feynman expansion, for example) is that the coefficients appearing in such an expansion, determined by cuts of loop amplitudes, are expressed in terms of {\it on-shell functions}---manifestly gauge invariant functions of observable states defining the theory. 

The main problem with the traditional approach of generalized unitarity, however, is that it is not {\it prescriptive}: it requires an arbitrary choice of the basis of integrands and sufficient computer power to solve the linear algebra problem of matching field theory on all cuts. These issues rapidly become computationally prohibitive. More importantly however, as with any such problem in linear algebra, the form of the solution that results depends strongly on the choice of basis. Some bases are better than others. 

This work is motivated by the desire to choose a `good' basis of loop integrands---one for which each term supports a unique, defining cut of field theory.\footnote{Of course, Cauchy's theorem does not allow a rational function to have only a single residue. In general, the integrands in our basis will contribute on many field theory cuts; however, the prescriptive nature is reflected in the fact that there is some cut (chosen from a `spanning set' of cuts  \cite{Bern:2015ooa}) for each integrand not shared by any other---thereby fixing the integrand's coefficient.} In such a basis, no linear algebra is required: the coefficient of each integrand is simply the corresponding field theory cut (a specific on-shell function). If each coefficient $c_k$ is a single on-shell function, we will say that the representation in (\ref{generalized_unitarity_intro}) is {\it prescriptive}; and we refer to the method followed to construct such a representation as {\it prescriptive unitarity}. This general strategy was introduced in \mbox{ref.\ \cite{Bourjaily:2013mma}} and used in \mbox{ref.\ \cite{Bourjaily:2015jna}} to construct closed-form representations of all two loop amplitudes in planar, maximally supersymmetric ($\mathcal{N}\!=\!4$) Yang-Mills theory (SYM). In the only-slightly schematic formalism used in this work, that result can be recast as follows,
\vspace{-5pt}\eq{\mathcal{A}_n^{L=2}=\bigger{\displaystyle\sum_{\raisebox{2pt}{\scalebox{0.6}{$\!\mathcal{L}$}}}}f_{\mathcal{L}}\fwbox{90pt}{\fig{-24.62pt}{1}{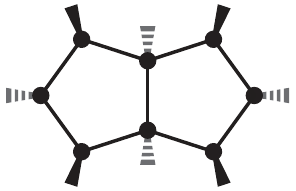}}\vspace{-5pt}\label{two_loop_amplitudes_intro}}
where each coefficient, $f_{\mathcal{L}}$, represents a single field theory residue. 

In this work, we describe how prescriptive unitarity can be applied to the case of planar SYM as our primary illustrative example. In particular, we show how this strategy can be used to construct an explicit, prescriptive representation for all $n$-point N$^k$MHV scattering amplitudes through three loops:
\vspace{-14pt}\eq{\mathcal{A}_n^{L=3}=\bigger{\displaystyle\sum_{\raisebox{2pt}{\scalebox{0.6}{$\!\mathcal{W}$}}}}f_{\mathcal{W}}\,\,\fwbox{100pt}{\fig{-47.3pt}{1}{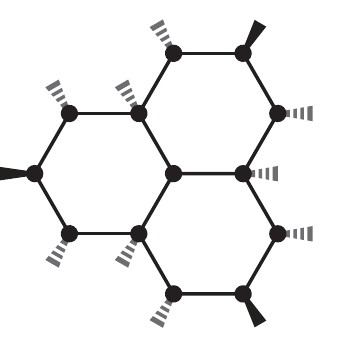}}\hspace{-10pt}\bigger{+}\hspace{5pt}\bigger{\displaystyle\sum_{\raisebox{2pt}{\scalebox{0.6}{$\!\mathcal{L}$}}}}f_{\mathcal{L}}\fwbox{130pt}{\fig{-34.76pt}{1}{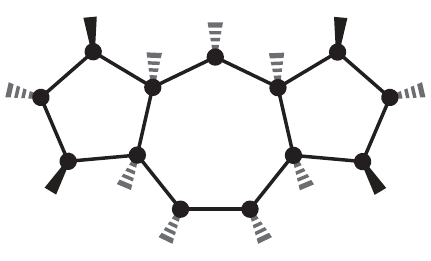}}\,,\label{general_three_loop_rep_intro}\vspace{-11.85pt}}
where every coefficient of every integral is a single, specific field-theory cut.

We should emphasize that our choice to apply these ideas to this particularly simple quantum field theory is motivated mostly by brevity in illustration; we expect that such representations exist in general. This optimism, however, requires testing through explicit construction---at higher orders of perturbation and for more general theories. As we will see below, even for this simple quantum field theory, the existence of a three loop prescriptive basis of integrands is rather non-trivial. Therefore, this work represents a concrete test that prescriptive representations exist.

This work is organized as follows. In \mbox{section \ref{sec:prescriptive_unitarity}} we review the basic ingredients of generalized unitarity and introduce the prescriptive unitarity approach to the (re)construction of loop amplitudes at the integrand-level. After briefly describing the cuts of loop amplitudes and integrands in \mbox{section \ref{subsec:generalized_unitarity_review}}, we discuss the traditional representation of one loop amplitudes via unitarity in \mbox{section \ref{subsec:one_loop_unitarity}}. In \mbox{section \ref{subsec:two_loop_unitarity}} we illustrate the prescriptive reformulation of unitarity, describing in detail how such a representation of all two loop amplitudes in planar SYM can be found. The representation we construct in \mbox{section \ref{subsec:two_loop_unitarity}} is different from that presented in \mbox{ref.\ \cite{Bourjaily:2015jna}}. This is both for the sake of conceptual clarity (in order to make it more analogous to our three loop result) and also for brevity. Most importantly, we have decided to ignore making the exponentiation of infrared divergences manifest at the integrand-level. Finally, in \mbox{section \ref{subsec:generalities_of_prescriptive_unitarity}} we outline how the prescriptive approach could be applied to more general quantum field theories. 
 
In \mbox{section \ref{sec:three_loop_result}} we use the prescriptive approach to construct a closed-form representation for all three loop amplitude integrands in planar SYM. The construction of the basis is described in \mbox{section \ref{subsec:three_loop_basis_summary}}, and the choices of cuts (and coefficients) involved in defining this basis are illustrated and described in \mbox{section \ref{subsec:exempli_cut_rules}}. The complete description of the terms appearing in our representation of three loop amplitudes is given in \mbox{Appendix \ref{appendix:details_three_loop_result}}. We conclude \mbox{section \ref{sec:three_loop_result}} with a general discussion of this representation in \mbox{section \ref{subsec:discussion_of_result}}. In \mbox{section \ref{sec:one_loop_revisited}} we revisit one loop (prescriptive) unitarity for theories with general power counting in four dimensions before outlining the prospects for future work in \mbox{section \ref{sec:conclusions}}.

Throughout this work, we have actively endeavored to keep our expressions free of any unnecessary reference to a particular choice of kinematical variables---namely, in the representation of loop integrands and their residues. Because of this, however, some of our work may appear unfamiliar even to the most expert of readers. We hope, however, that the examples and illustrations used in our review are sufficiently clear to make both our result and the more general strategy more accessible.

\newpage
\vspace{-2pt}\section{From Generalized to {\it Prescriptive} Unitarity}\label{sec:prescriptive_unitarity}\vspace{-2pt}

The basic idea of generalized unitarity is very simple: because Feynman diagrams are rational functions prior to loop integration, the loop {\it integrands} of arbitrary scattering amplitudes are rational functions of the external and internal momenta; being rational functions, they are expandable into a complete basis of functions, with coefficients determined by residues (or `poles'). Schematically, suppose $\{\mathcal{I}_k\}$ forms a complete basis of $L$ loop integrands (appropriate for a particular  field theory), then an arbitrary scattering amplitude integrand, $\mathcal{A}_n^{L}$, can be represented:
\vspace{-2.5pt}\eq{\mathcal{A}_n^L=\sum_{k}c_k\,\mathcal{I}_k\,.\label{schematic_unitarity_expansion}\vspace{-4pt}}
The coefficients in this expansion, $c_k$, are determined by the criterion that the right hand side matches field theory on all residues (of arbitrary co-dimension). Depending on the choice of basis $\{\mathcal{I}_k\}$, the coefficients $c_k$ in (\ref{schematic_unitarity_expansion}) may be individual residues or arbitrarily complicated linear combinations thereof. {\it A representation of the form} (\ref{schematic_unitarity_expansion}) {\it will be called {\bf prescriptive} if every coefficient $c_k$ is a single field-theory cut.}

Before describing representations of the form (\ref{schematic_unitarity_expansion}) in more detail, it is worth taking a moment to explain why such a representation would be advantageous. The primary reasons are two-fold. First, although integrated amplitudes can be horrendously complicated transcendental functions, the residues of amplitude integrands are always gauge-invariant algebraic functions of physical observables constructed from tree amplitudes (and enjoying all the symmetries of tree-level $S$-matrices). Thus, regardless of the complex linear combinations of cuts that may appear in the coefficients upon solving the constraints, the terms involved are (relatively) easy to compute, with complexity similar to that of tree amplitudes. Secondly, once a choice of basis is fixed, each integral may be integrated and tabulated irrespective of any particular quantum field theory (provided the basis is sufficiently general). And because loop integration remains considerably harder than integrand construction, it is extremely convenient to reuse integrals for many different computations or reduce them to a smaller set of master integrals, see {\it e.g.}\ \cite{Smirnov:2006ry,Smirnov:2012gma,Henn:2014qga,Zhang:2016kfo} and references therein. While we are not using integral reduction in this work, we are simplifying the work required to find integrand-level representations, after which integration-by-parts identities could be exploited. For related work trying to identify certain master integrands that are nonzero upon integration, see {\it e.g.}\ \cite{Kosower:2011ty,Gluza:2010ws,Johansson:2015ava,Ita:2015tya}.

\vspace{-0pt}\subsection{The Generalized Unitarity Approach to Integrand Construction}\label{subsec:generalized_unitarity_review}\vspace{-2pt}

In order to be more precise about the ingredients involved in representing an amplitude according to (\ref{schematic_unitarity_expansion}), it will be useful to define some basic notation and conventions. The kinds of integrands in which we will be interested consist of some number of ordinary Feynman propagators, of the form $1/(\ell_i\mi p)^2$ where $\ell_i$ is one of the $L$ loop momenta, and $p$ is some combination of external/internal momenta, with numerators given as polynomials in $\ell_i$ constructed out of Lorentz invariants. The degree of the numerator polynomials is dictated by the field theory in question.

We will have more to say about the numerators soon, but for now let us introduce the notation used for the denominators. The integrals in which we are interested will have denominators corresponding to some scalar Feynman diagram---with each factor of the form $(\ell\mi p)^2$. For reasons of notational simplicity (and kinematic agnosticism), we will write these propagators according to the {\it edges} of a Feynman graph: using `$\x{\ell}{a}$' to denote the squared momentum flowing through edge `$a$' of the graph. For example, a one loop integrand involving five propagators (a `pentagon') would be written:
\vspace{-6pt}\eq{\fig{-34.76pt}{1.36}{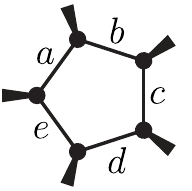}\bigger{\equiv}\;\frac{N(\ell)}{\x{\ell}{a}\x{\ell}{b}\x{\ell}{c}\x{\ell}{d}\x{\ell}{e}}\,,\label{edge_labeled_pentagon}\vspace{-6pt}}
where $N(\ell)$ is some polynomial in $\ell$. For plane graphs, every edge can be unambiguously labelled by the Poincar\'{e}-dual faces which they connect; and each face can unambiguously labelled by external or internal momenta. Thus, we may use the same labels for edges as for external legs or internal momenta. Our convention will be that $\x{\ell}{a}$ denotes the `external' propagator\footnote{By {\it external}, we simply mean the Feynman propagator of a loop momentum which is on the exterior of the graph.} preceding the external leg $a$ clockwise around the graph, and $\x{\ell_i}{\ell_j}$ denotes an `internal' propagator between loop momenta $\ell_i,\ell_j$. Thus, the edges in (\ref{edge_labeled_pentagon}) would correspond to a graph with external legs labelled:
\eq{\fig{-34.76pt}{1.36}{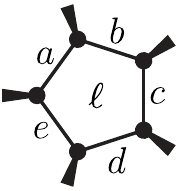}\!\!\bigger{\Leftrightarrow}\hspace{5pt}\fig{-34.76pt}{1.36}{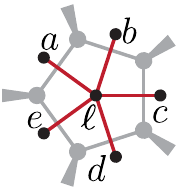}\!\!\bigger{\Leftrightarrow}\hspace{2.5pt}\fig{-34.76pt}{1.36}{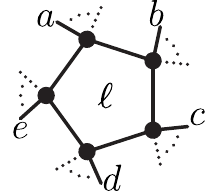}\label{edge_leg_labeling_example}}
Throughout this work, we will label leg ranges spanning an arbitrary (but non-empty) length using the notation of the figure on the left in (\ref{edge_leg_labeling_example}). Later on, we will also make use of dashed wedges to indicate ranges of legs that may possibly be empty. 

Although we have endeavored to keep our formulae kinematically agnostic, it is worth mentioning how natural this notation is when integrands are expressed in dual momentum coordinates. These coordinates are linearly related to ordinary momenta, but make momentum conservation (and translational invariance) manifest. Specifically, we could choose to express the $a^{\text{th}}$ external momentum as $p_{a}\!\equiv\!x_{a+1}\!\mi x_a$ (with $x_{n+1}\!\simeq\! x_1$ being understood); the points $x_a$ are called `dual momentum coordinates'. Differences in these coordinates then represent sums of consecutive momenta, $x_b\mi x_a\!=\!p_a\!\pl p_{a+1}\!\pl\!\ldots\!\pl p_{b-1}$, so that:
\eq{\x{a}{b}=\x{b}{a}\equiv(x_b\mi x_a)^2=(p_a+p_{a+1}+\ldots+p_{b-1})^2\,.\label{dual_coordinates_prop}}
For planar integrands in these coordinates, each loop momentum would be assigned a dual coordinate $x_{\ell_i}$, so that all propagators explicitly take the form $\x{\ell}{a}\!\equiv\!(x_{\ell}\mi x_a)^2$ or $\x{\ell_i}{\ell_j}\!\equiv\!(x_{\ell_i}\mi x_{\ell_j})^2$.

\vspace{-0pt}\subsubsection{On-Shell Functions: the Cuts of Loop Amplitudes}\label{subsubsec:on_shell_functions_and_loop_integrands}\vspace{-2pt}

Given a basis of loop integrands, the criterion used to fix the coefficients $c_k$ in (\ref{schematic_unitarity_expansion}) is that the residues of the right hand side match field theory. We presume the reader understands how to compute the (multidimensional) residues (see {\it e.g.}\ \cite{GriffithsHarris}) of explicit, rational integrands. For a precise definition of cut integrals, see the recent work \cite{Abreu:2017ptx}. The residues of field theory amplitudes should also be familiar, but are worth reviewing---if only to clarify notation that will be used throughout this work. 

The residues of a scattering amplitude are those functions obtained from an off-shell ({\it e.g.}\ Feynman) loop integrand by setting some subset of internal particles on-shell. If starting from the Feynman expansion, it is not hard to see that the set of all Feynman graphs sharing some subset of internal propagators will span entire lower loop amplitudes at the vertices.\footnote{This is strictly true for theories with any amount of supersymmetry; for more general theories, this statement requires at least two propagators be cut.} Thus, the residues of loop amplitudes correspond to graphs with amplitudes at each vertex separated by on-shell, internal states. Functions corresponding to such graphs are called {\it on-shell functions}, and they have played a key role in many of the recent developments in our understanding of scattering amplitudes. 

On-shell functions can be defined (and computed) in many ways---using many kinematical choices which may simplify their form for particular theories (in particular dimensions, etc.). Even though they have been most prominently featured in the realm of supersymmetric theories \cite{ArkaniHamed:2012nw,Arkani-Hamed:2014bca,Franco:2015rma,Frassek:2016wlg,Heslop:2016plj,Herrmann:2016qea} (see {\it e.g.} \cite{Benincasa:2015zna,Benincasa:2016awv} for some exceptions), they can generally be defined from first principles without reference to (off-shell) loop integrands in any quantum field theory. When represented as a graph $\Gamma$ of amplitudes at vertices indexed by $v$, connected by edges indexed by $i$ (representing on-shell, but internal physical states), the corresponding on-shell function $f_\Gamma$ can be defined simply by:
\eq{f_\Gamma\equiv\prod_{i}\Big(\!\!\sum_{\text{states}}\int\!\!d^{d-1}\text{LIPS}_i\!\Big)\!\!\prod_{v}\mathcal{A}_v\,.\label{general_on_shell_function_definition}}
This definition follows immediately from locality and unitarity. In (\ref{general_on_shell_function_definition}), `$d\text{LIPS}$' denotes the measure over the `Lorentz invariant phase space' of each on-shell, internal particle, and the summation over `states' means over all non-kinematical quantum labels distinguishing particles in the theory---helicity, colour, etc. We hope that the reader appreciates that (\ref{general_on_shell_function_definition}) has been written in so as to make clear that these objects are definable in arbitrary numbers of dimensions. 

For many of the on-shell functions important to this work, the phase space integrations in (\ref{general_on_shell_function_definition}) are not entirely localized by the momentum conservation at the vertices; when this happens, $f_\Gamma$ becomes an (unspecified) integral over on-shell degrees of freedom. The integrand of $f_\Gamma$ is thus some generally algebraic (often rational) function of both external and internal, always on-shell degrees of freedom. 

As discussed above, on-shell functions may be equivalently defined as the iterated residues of off-shell loop amplitudes obtained by putting each edge in the diagram on-shell. On-shell functions defined in this way (as residues of loop amplitudes) appeared first historically in the context of generalized unitarity. While we prefer the first-principles definition (\ref{general_on_shell_function_definition}), this historical view is useful to bear in mind. For example, considering on-shell functions as residues makes it easy to count how many `internal' degrees of freedom exist for a given diagram: an $L$-loop diagram with $n_I$ internal edges corresponds to a co-dimension $n_I$ residue of a $(d\!\times\!L)$-dimensional form---resulting in a function of $(d\!\times\!L\mi n_I)$ remaining degrees of freedom. In this work we will mostly be concerned with $d\!=\!4$-dimensional quantum field theories. 

Among the most important of all on-shell functions is the `unitarity' cut:
\vspace{-7pt}\eq{\fig{-17.45pt}{1.36}{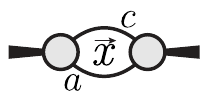}\equiv\frac{1}{x_1x_2}\mathcal{A}_L(\ell^a(\vec{x}),p_a,\ldots,\mi\ell^c(\vec{x}))\mathcal{A}_R(\ell^c(\vec{x}),p_c,\ldots,\mi\ell^a(\vec{x}))\,,\label{bubble_definition}\vspace{-7pt}}
where $\ell^a,\ell^c$ are the on-shell momenta flowing through the corresponding edges---obtained as solutions to $\x{\ell}{a}\!=\!\x{\ell}{c}\!=\!0$ and expressed in terms of the 2 remaining degrees of freedom, $\vec{x}\!\equiv\!(x_1,x_2)$. From this trivial (if essential) starting point, all one-loop on-shell functions can be obtained by taking iterated residues---{\it e.g.},
\vspace{-10pt}\eq{\Res_{\x{\ell}{b}=0}\left(\raisebox{20pt}{$\!\!$}\right.\hspace{-5pt}\fig{-34.76pt}{1.36}{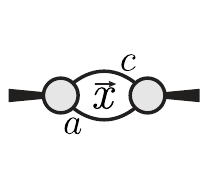}\hspace{-5pt}\left.\raisebox{20pt}{$\!\!$}\right)\,{=}\;\fig{-34.76pt}{1.36}{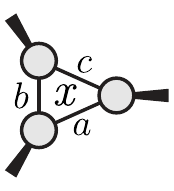}\,.\label{bubble_to_triangle}\vspace{-10pt}}
Here, the pole in $\x{\ell}{b}$ must arise from one of the amplitudes in (\ref{bubble_definition}), and the resulting on-shell function has one internal degree of freedom---now denoted simply by `$x$'. If one further residue is taken, the result is an on-shell function without any internal degrees of freedom---a `leading singularity' (in four dimensions):
\vspace{-10pt}\eq{\Res_{\x{\ell}{d}=0}\left(\raisebox{32pt}{$\!$}\right.\fig{-34.76pt}{1.36}{triangle_cut}\left.\raisebox{32pt}{$\!\!$}\right)\,\,=\fig{-34.76pt}{1.36}{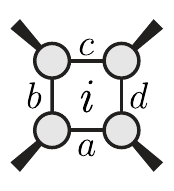}\equiv\, f_{abcd}^i\,.\label{triangle_to_box}\vspace{-10pt}}
Although the on-shell function on the right-hand side (\ref{triangle_to_box}) has no internal degrees of freedom, it carries a label `$i$' to distinguish between the two ``quad-cut'' solutions, denoted `$Q^i_{abcd}$', to the four simultaneous on-shell conditions:
\vspace{-5pt}\eq{\x{\ell^*}{a}=\x{\ell^*}{b}=\x{\ell^*}{c}=\x{\ell^*}{d}=0\quad\text{for}\quad\ell^*\!\in\!\{Q^1_{abcd},Q^2_{abcd}\}\,.\label{quad_cuts_defined}\vspace{-5pt}}
The reason for making such a distinction is that the on-shell functions for the two solutions (which are now basically just products of tree amplitudes, evaluated on $\ell^*$) are related by parity but are generally distinct: most of the time, $f^1_{abcd}\!\neq\!f^2_{abcd}$. The details of their functional form will not be important to us. However, that \emph{there are \emph{two} solutions} to cutting four propagators in four dimensions and that \emph{the resulting on-shell functions are generally distinct} will be very useful facts to bear in mind. 

Although we will use (planar) SYM as the primary example throughout this work, we would like to emphasize that the precise form taken for the on-shell functions in this theory will play essentially no role whatsoever. Indeed, most of our analysis would be valid for any particular (four-dimensional) quantum field theory. If every on-shell function in this work were reinterpreted as those of non-supersymmetric Yang-Mills, for example, virtually all of our results would remain valid: our formulae would represent important and correct---merely {\it incomplete}---contributions to loop amplitude integrands in pure Yang-Mills. For those readers interested in a more concrete understanding of the on-shell functions used in this work for planar SYM, we refer the reader to more thorough discussions in the literature (see {\it e.g.}\ \cite{ArkaniHamed:2012nw,ArkaniHamed:book} or the appendices of \cite{Bourjaily:2015jna}), and to the computer packages described in \mbox{refs.\ \cite{Bourjaily:2010wh,Bourjaily:2012gy,Dixon:2010ik}}.

Before moving on, we should clarify some of the terminology that is often used to describe on-shell functions. For this work, we consider `residues' and `cuts' to be interchangeable. Residues with maximal co-dimension (which may involve $d\!\times\!L$ internal propagators, or simply $d\!\times\!L$ cut conditions among fewer propagators) have no internal degrees of freedom. These on-shell functions will play an important role for us; they are called `{\it leading singularities}' \cite{Cachazo:2008vp}. Residues for which the number of cut conditions exceeds the number of internal propagators are called {\it composite}. (Composite residues, however, will not be very important to our present work.) Residues which depend on some number of internal degrees of freedom---such as the unitarity cut (\ref{bubble_definition})---may occasionally be called `sub-leading' singularities. 

Closely related to (sub-)leading singularities are the so-called `maximal cuts' \cite{Bern:2007ct} (see also \cite{Bosma:2017ens}). Maximal cuts are those residues which cut the maximum number of internal propagators of an amplitude; this number depends on multiplicity, but can be substantially less than \mbox{$d\!\times\!L$}. As such, maximal cuts often correspond to what we call sub-leading singularities---which could potentially be a source of confusion. We choose not to use the language of maximal- and (next-to)$^k$-maximal cuts, however, because the counting of the number of internal degrees of freedom of a given residue is much more important to us than the number of propagators involved (or the multiplicity of an amplitude).

\vspace{-0pt}\subsection{Generalized Unitarity at One Loop}\label{subsec:one_loop_unitarity}\vspace{-2pt}

In this subsection, we briefly review (traditional) generalized unitarity at one loop \cite{Bern:1994zx,Bern:1994cg,Britto:2004nc}. This will provide a convenient excuse to introduce some essential aspects about integrand reduction, and illustrate the differences with the {\it prescriptive} approach we describe here. For the sake of clarity and concreteness, let us restrict ourselves to four-dimensional quantum field theories. 

Let us review some classical results about integrand reduction in four dimensions. The space of squared propagators is easily seen to be six-dimensional: any such factor can always be expanded into a ``na\"{i}ve'' basis of Lorentz-invariant monomials:
\vspace{-2.5pt}\eq{\x{\ell}{Y}\equiv(\ell-p_Y)^2\in\text{span}\underset{\text{``na\"{i}ve'' basis for inverse propagators}}{\big\{1,\ell\!\cdot\!k_1,\ell\!\cdot\!k_2,\ell\!\cdot\!k_3,\ell\!\cdot\!k_4,\ell^2\big\}}\,.\label{naive_basis}\vspace{-2.5pt}}
Here, $k_i$ represent any spanning set of four-dimensional momenta. There are various ways to make this counting manifest---for example, using momentum twistors \cite{Hodges:2009hk} or projective coordinates (see {\it e.g.}\ \cite{Weinberg:2010fx}); but for our purposes, the obvious counting in (\ref{naive_basis}) will suffice. We restrict our discussion to general kinematics and will not exploit any linear dependencies that may arise for low-point kinematics (when $n\!\leq\!d$), \cite{Passarino:1978jh,Ellis:2011cr}.

The fact that inverse propagators (in four dimensions) span a six-dimensional space has some immediate consequences independent of any particular quantum field theory. One fairly trivial consequence of (\ref{naive_basis}) is that any polynomial of degree less than three in loop momenta can be expressed alternatively in a six-dimensional space of inverse propagators---{\it including the identity polynomial}. In particular, this means that---regardless of the (field-theory-determined) power counting of numerators, any integrand with six or more propagators can be expanded in terms of those involving five or fewer: simply choose six of the inverse propagators to expand `$1$' in the numerator, resulting in terms with strictly fewer propagators. This can be done recursively for any integral until all terms have at most five propagators. This reduction procedure was first described by Passarino and Veltman in \mbox{ref.\ \cite{Passarino:1978jh}}. This generalizes trivially at higher loop orders to imply that any integrand involving six or more {\it external} propagators is reducible into those involving five or fewer. (We will see below that this statement can be strengthened for planar theories.)

Let us now discuss the forms taken for loop-dependent numerators of four-dimensional integrands. As we have seen, all Lorentz-invariant monomials can be expanded in a six-dimensional basis of inverse propagators; this implies that we may without any loss of generality consider only numerators constructed as products of inverse propagators. It is not hard to see that the space of $r$ powers of inverse propagators spans a space whose rank is given by:
\vspace{-2.5pt}\eq{\text{rank of $r$-fold products of inverse propagators: } \binom{r+3}{4}+\binom{r+4}{4}\,.\label{dimension_counting_by_power}\vspace{-2.5pt}}
This counting follows from the fact that these functions correspond to symmetric, traceless products of $\mathbf{6}$'s of $SO_6$. (For this work, the most important instances of (\ref{dimension_counting_by_power}) are for $r\!=\!1,\!2$---polynomials with $6$ and $20$ degrees of freedom, respectively.)

From the discussion above, it should be clear that all one loop amplitude integrands in any four-dimensional quantum field theory can be expanded in terms of integrals involving at most five propagators. In order to fully specify a basis of integrals in (\ref{schematic_unitarity_expansion}), however, we must also know the highest power of loop momentum that can appear in numerators. This is determined by the ultraviolet behavior of the theory in question.

For the sake of illustration, let us consider the case of (maximally) supersymmetric Yang-Mills theory (SYM). Amplitudes in this theory scale as $\sim\!1/(\ell^2)^4$ at large loop momentum, so that any integral involving five propagators as in (\ref{edge_labeled_pentagon}) should have a numerator of the form $\x{\ell}{Y}$---a six-dimensional space of possible numerators. This suggests that for any $n$-point amplitude we can construct a complete basis of integrands in terms of $\binom{n}{5}$ pentagon integrals with $6$ degrees of freedom each, and $\binom{n}{4}$ box integrals with $1$ degree of freedom each (their overall normalization). Such a basis of integrands would indeed be complete---but considerably {\it over}-complete. 

One way to see that this set of integrals forms an over-complete basis is to choose a convenient basis for the numerators of each pentagon integral. Consider the pentagon drawn in (\ref{edge_labeled_pentagon}). Its numerator has $6$ degrees of freedom; a natural choice of basis for these would involve the $5$ relevant inverse propagators, together with the dual of these $5$---generated by the six-dimensional $\epsilon$-tensor:\footnote{The tensor $\epsilon(\ell,\!a,\!b,\!c,\!d,\!e)$ can be expanded in terms of inverse propagators (involving additional (complex) momenta), but its actual form will not be important for us here.} 
\eq{\x{\ell}{Y}\equiv(\ell-p_Y)^2\in\text{span}\underset{\text{``parity'' basis for inverse propagators---five external}}{\big\{\overset{\text{{\color{hblue}non-contact}}}{{\color{hblue}\epsilon(\ell,a,b,c,d,e)}},\overset{\text{{\color{hred}contact}}}{{\color{hred}\x{\ell}{a}},{\color{hred}\x{\ell}{b}},{\color{hred}\x{\ell}{c}},{\color{hred}\x{\ell}{d}},{\color{hred}\x{\ell}{e}}}\big\}}\,.\label{parity_basis_for_props}}
In this basis, $5$ of the $6$ degrees of freedom of each pentagon directly give rise to box integrals---without any loop dependence in their numerators. These are called `contact terms' of the original pentagon; and we see that the $6$ degrees of freedom of any pentagon integral cleanly separate into ${\color{hred}5}$ contact terms, and only ${\color{hblue}1}$ non-contact degree of freedom. This is responsible for (most) of the redundancy in our na\"{i}ve basis of $\binom{n}{5}$ pentagons with general numerators and $\binom{n}{4}$ scalar boxes with loop-independent numerators. 

The decomposition of pentagon numerators according to (\ref{parity_basis_for_props}) is called the ``parity'' basis because it naturally separates all integrands into {\it scalar} box integrals (symmetric under parity), and parity-odd pentagon integrals. Because the $\epsilon$-tensor in (\ref{parity_basis_for_props}) changes sign under parity, these integrals vanish upon integration (on the parity-even Feynman contour of loop momenta). As such, they are irrelevant to {\it integrated} amplitudes and their role in representing loop integrands is consequently, often neglected.

Before we can discuss how amplitudes in SYM are represented according to (\ref{schematic_unitarity_expansion}) using this basis of integrals, we must first observe that it is {\it still} over-complete! To correctly count the total degrees of freedom required to expand any integral, imagine first combining all terms over a common denominator built from all $n$ propagators. The power counting discussed above implies that the amplitude must have $(n\mi4)$ powers of inverse propagators in the numerator, implying a total number of degrees of freedom given by (\ref{dimension_counting_by_power}) with $r\!=\!(n\mi4)$. And so, while parity-odd pentagons and scalar boxes do form a basis, they represent $\binom{n}{5}\pl\binom{n}{4}$ degrees of freedom, which exceeds the correct number, $\binom{n}{4}\pl\binom{n-1}{4}$, for $n\!\geq\!6$. Indeed, we can see from this counting that the parity-odd pentagons satisfy $\binom{n-1}{5}$ integrand-level relations, which {\it must} be eliminated in order for us to uniquely fix the coefficients of the chosen, independent subset of pentagons. 

The upshot of the preceding discussion is that we know that there exists a representation of one loop integrands in SYM of the form:
\vspace{-2.5pt}\eq{\mathcal{A}_n=\sum c_{abcde}\,\mathcal{I}_{abcde}+\sum c_{abcd}\,\mathcal{I}_{abcd}\,,\label{one_loop_integrand_unitarity}\vspace{-2.5pt}}
where the first terms include some choice of independent parity-odd pentagon integrals. This choice obviously affects the complexity of the coefficients that arise, but has no impact on the coefficients of the scalar boxes---for the simple reason that only these terms survive upon integration, and therefore cannot depend on the choice of basis for the parity-odd pentagons. Thus, if we were only concerned with integrated amplitudes, the representation simplifies considerably:
\vspace{-2.5pt}\eq{\int\!\!d^4\ell\,\,\mathcal{A}_n=\sum c_{abcd}\,\int\!\!d^4\ell\,\,\mathcal{I}_{abcd}\,.\label{integrated_one_loop_unitarity}\vspace{-2.5pt}}
Because this expression is independent of the choice of basis for the parity-odd pentagons, it certainly appears prescriptive. Indeed, the coefficients of the scalar box integrals are (deceptively) simple:
\vspace{-15pt}\eq{c_{abcd}=\sum_{i=1,2}f_{abcd}^i\,\quad\mathrm{where}\quad f_{abcd}^i\equiv\!\! \fig{-34.76pt}{1.36}{box_cut}\label{scalar_box_coefficients}\vspace{-10pt}}
where $f_{abcd}^i$ are on-shell functions corresponding to cutting the obvious four propagators, which are summed over the two (parity-conjugate) leading singularities with the topology of a given box.

The fact that the coefficients of the scalar boxes take this simple form is not hard to prove by considering the co-dimension four residues of the amplitudes and box integrals in the basis. But this simplicity hides a great deal of underlying structure that is easily overlooked. For example, not all co-dimension four residues of amplitudes involve four propagators: there are also the so-called `composite' residues involving only three propagators separated by two massless legs:
\vspace{-14pt}\eq{\fwbox{68pt}{\fig{-34.76pt}{1.36}{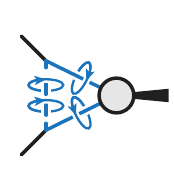}}=\fwbox{30pt}{\fig{-34.76pt}{1.36}{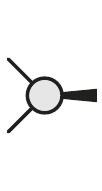}}\qquad\text{supported on box integrals via}\quad\fig{-34.76pt}{1.36}{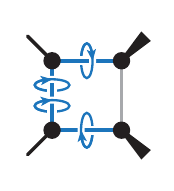}\label{one_loop_composites}\vspace{-14pt}}
These residues are supported where the loop momentum becomes both soft and collinear (to some external leg), and exist within the range of the Feynman contour (for $\ell\!\in\!\mathbb{R}^{3,1}$); as such, they are precisely responsible for the infrared divergences of loop amplitudes. The fact that the representation  (\ref{integrated_one_loop_unitarity}) gets these non-manifestly-matched residues of field theory correct follows from the completeness of our basis and the fact that parity-odd pentagons always vanish on these parity-even residues. But as indicated in (\ref{one_loop_composites}), these residues of field theory are simply tree amplitudes; as such, the fact that the box expansion (\ref{integrated_one_loop_unitarity}) reproduces these cuts is how the tree-level BCFW recursion relations were originally discovered \cite{BCF} (only later proven in \mbox{ref.\ \cite{BCFW}}). In what follows, we will not make use of composite residues in our work mostly because they exist only for integrals involving massless legs---which would require us to deal with the various cases of possible leg distributions differently. See the end of \mbox{section \ref{subsec:two_loop_unitarity}} for a more thorough discussion of the advantages and disadvantages of making these residues (responsible for infrared divergences) manifest. 

What we have described so far is a more thorough version of how generalized unitarity is usually described at one loop. The representation (\ref{one_loop_integrand_unitarity}) does not meet our requirement for being {\it prescriptive} for the simple reason that the coefficients are not individual residues. Despite the fact that the integral level statement in (\ref{integrated_one_loop_unitarity}) is very nearly prescriptive, there is no way to avoid choosing a basis of parity-odd pentagons in (\ref{one_loop_integrand_unitarity}), and the mess of linear algebra resulting in their coefficients. 

This story can in fact be recast in a prescriptive way, but doing so requires several complications unnecessary beyond one loop (if we insist on maintaining the manifest power counting of SYM). After describing the prescriptive approach to amplitudes at two and three loops, it will be much easier to understand prescriptive unitarity at one loop. Thus, we postpone a more general discussion of one loop prescriptive unitarity until \mbox{section \ref{sec:one_loop_revisited}}, where we will see that weakening the limits on the power counting of the theory will allow us to better describe SYM at one loop.

\newpage
\vspace{-6pt}\subsection{{\it Prescriptive} Unitarity at Two Loops ({Redux})}\label{subsec:two_loop_unitarity}\vspace{-6pt}

In the past, increasing the loop order or the number of legs often led to computational challenges. Some of the early results started with the computation of integrands for fixed number of legs, \cite{Bern:1997nh}, which were later extended to arbitrary multiplicity, \cite{Vergu:2009tu,ArkaniHamed:2010kv,ArkaniHamed:2010gh,Bourjaily:2015jna}. In pure Yang-Mills, explicit results for all plus amplitudes up to six points are also available \cite{Badger:2013gxa,Badger:2015lda,Gehrmann:2015bfy,Dunbar:2016aux,Dunbar:2016gjb}, see also work on numerical unitarity methods at two loops \cite{Abreu:2017xsl}. 

Surprisingly enough, matching two loop amplitudes in planar SYM at the integrand-level according to unitarity (even prescriptively) turns out to be simpler than at one loop. To see this, let us first describe the analog to Passarino-Veltman reduction \cite{Passarino:1978jh} relevant to amplitudes in (planar) SYM. Without any loss of generality, we may consider all integrals to include at least one internal propagator (multiplying by $\x{\ell_1}{\ell_2}/\x{\ell_1}{\ell_2}$ if necessary). 

Power counting now requires that any integrand in a purported basis must involve at least three external propagators per loop (four propagators total per loop). How many external propagators are allowed before integrand reduction implies dependencies? For reasons that we will soon demonstrate, it turns out the answer is four, resulting in a (possibly over-)complete basis from the following topologies:
\vspace{-2.5pt}\eq{\left\{\fwbox{60pt}{\fig{-24.6pt}{1}{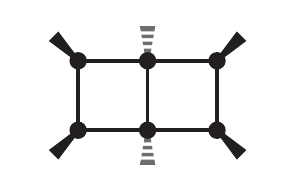}},\fwbox{85pt}{\fig{-24.6pt}{1}{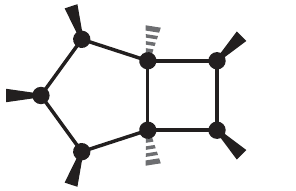}}\hspace{-10pt},\fwbox{85pt}{\fig{-24.6pt}{1}{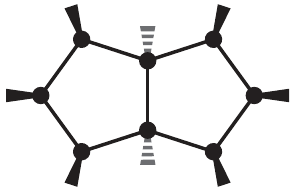}}\right\}\bigger{\subset}\left\{\fwbox{80pt}{\fig{-24.6pt}{1}{two_loop_ladder_general}}\right\}\,.\label{two_loop_topologies}\vspace{-2.5pt}}
The first of these topologies has no loop dependence and only $1$ degree of freedom in the numerator, the second has $6$ degrees of freedom, and the third has $6\!\times\!6$. In order for us to see that no integrands involving more external propagators are required, it will be helpful to first describe the degrees of freedom of these integrands. 

Consider first the `pentabox' integral---the second topology in (\ref{two_loop_topologies}). This integral's numerator must involve a single inverse propagator. It will be useful to describe these $6$ degrees of freedom in terms of contact/non-contact parts. Obviously, the contact terms are captured by the four relevant inverse propagators, leaving two non-contact degrees of freedom. These orthogonal degrees of freedom are naturally captured by two quad-cuts---the points in loop momentum space determined by putting the four external propagators on shell. Denoting the four external propagators of the pentagon-part of the pentabox integral by $\{\x{\ell}{a},\ldots,\x{\ell}{d}\}$, a natural basis for numerators would be given by:
\vspace{-2.5pt}\eq{\x{\ell}{Y}\equiv(\ell-p_Y)^2\in\text{span}\underset{\text{``chiral pentagon'' basis for inverse propagators---four external}}{\big\{\overset{\text{{\color{hblue}non-contact}}}{{\color{hblue}\x{\ell}{Q_{abcd}^1}},{\color{hblue}\x{\ell}{Q_{abcd}^2}}},\overset{\text{{\color{hred}contact}}}{{\color{hred}\x{\ell}{a}},{\color{hred}\x{\ell}{b}},{\color{hred}\x{\ell}{c}},{\color{hred}\x{\ell}{d}}}\big\}}\,.\label{chiral_pentagon_basis}\vspace{-2.5pt}}
Thus, the general numerator of a pentabox can be described as consisting of exactly ${\color{hblue}2}$ non-contact degrees of freedom, and ${\color{hred}4}$ contact terms---each having the topology of a double-box (the first picture in (\ref{two_loop_topologies})). 

Using the same basis for inverse propagators for each pentagon, it is easy to see that the $6^2$ degrees of freedom of a double-pentagon---the last topology in (\ref{two_loop_topologies})---can be decomposed according to $6^2\!=\!({\color{hblue}2}\pl{\color{hred}4})^2$. Although we could envision all the topologies in (\ref{two_loop_topologies}) as arising as contact terms of the double-pentagon, it turns out to be much smarter to discuss each topology as being defined {\it modulo} its contact-term degrees of freedom. Thus, a double-box has a single degree of freedom; a pentabox has 2 degrees of freedom (modulo contact terms); and a double-pentagon has 4 degrees of freedom (modulo contact terms). In what follows, we will define our basis using {\it all} the degrees of freedom for each topology (which would seem like a very over-complete basis), with an important role played by the {non-contact} degrees of freedom of each. Thus, we may reformulate our basis according to,
\vspace{-2.5pt}\eq{\left\{\fwbox{60pt}{\fig{-24.6pt}{1}{two_loop_ladder_1}},\fwbox{85pt}{\fig{-24.6pt}{1}{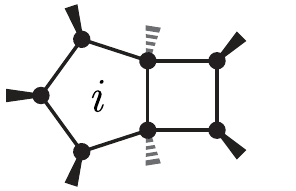}}\hspace{-10pt},\fwbox{85pt}{\fig{-24.6pt}{1}{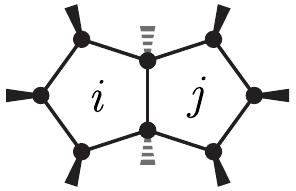}}\right\}\,,\label{two_loop_ints}\vspace{-2.5pt}}
where an index $i,j\!\in\!\{1,2\}$ denotes the non-contact degrees of freedom for each term. 

We are now ready to see that any integrand involving more than four external propagators for either loop momentum is reducible into the topologies given in (\ref{two_loop_ints}). Suppose that one of the loops involved five external propagators. Its $20$ degrees of freedom could then be spanned by $\binom{5}{1}$ single contact terms with $2$ (non-contact) degrees of freedom each, and $\binom{5}{2}$ double-contact terms with $1$ degree of freedom each (their  normalization). Thus, any integrand involving more than four external propagators is expandable into those in (\ref{two_loop_ints}). 

Before moving on, we should be clear that our present basis of two loop integrands (relevant to planar SYM) in (\ref{two_loop_ints}) is certainly over-complete. This is because, for example, while we consider there to be two pentaboxes (indexed by $i$, counting the non-contact terms), we are going to allow them to be defined by {\it all} their $6$ degrees of freedom of a general numerator consistent with power counting---{\it including their contact terms}. We will see below that these integrands, as they appear in the basis for our prescriptive representation, will have rules to specify all six of their degrees of freedom. This may seem to be a rather poor choice of basis, it being initially over-complete; however, these additional degrees of freedom will be critical to allowing us to construct a {\it strictly diagonal} basis for cuts---a basis of integrands for which each term matches a specific field theory residue unique to that integrand. Once such a basis has been constructed, its non-over-completeness is guaranteed.

\newpage
\vspace{-6pt}\subsubsection{Choosing a {\it Diagonalized} Basis of Integrands/Cuts}\label{subsubsec:diagonal_basis_of_cuts}\vspace{-4pt}

Let us now describe how to fully determine each integrand in our basis (\ref{two_loop_ints}) according to field theory cuts. The first of the integrals, the double-boxes, are the simplest but arguably the least trivial. They are simple because each double-box has only a single degree of freedom, and so we need only determine its normalization; they are the least trivial because they do not have (in the general case) any residues with maximal co-dimension. (When one or more of the middle leg ranges are empty, the integrals do have support on maximal co-dimension residues, but we choose here to ignore this potential simplicity in favor of a more generally valid approach.)

The fact that double-box integrals do not generally support residues with maximal co-dimension is not in fact very problematic: we merely need to match field theory on a less-than-maximal co-dimension residue. For example, let us choose to consider the co-dimension six residue of the integral that puts all six of the external propagators on-shell. We may parameterize the two-dimensional space of loop momenta along this residue by `$(x,y)$'---one parameter per loop. The residue of the six propagators is easy to take: it produces a simple Jacobian\footnote{This Jacobian appears on both sides of (\ref{schematic_unitarity_expansion}), and so is not actually relevant to the coefficients.}, together with the internal propagator, $1/\x{\ell_1(x)}{\ell_2(y)}$, left as a function of the remaining degrees of freedom:
\vspace{-2.5pt}\eq{\fwbox{60pt}{\fig{-24.6pt}{1}{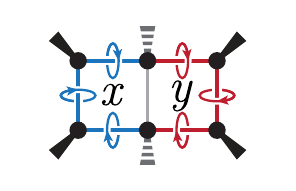}}\vspace{-2.5pt}}

The corresponding residue of field theory is similarly easy to evaluate, it also being a function of two internal degrees of freedom. We will represent this as:
\vspace{-2.5pt}\eq{\fwbox{60pt}{\fig{-24.6pt}{1}{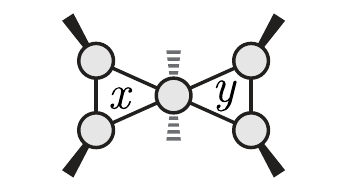}}\label{two_loop_cut_1}\vspace{-2.5pt}}
A closed formula for this on-shell function (expressed using momentum twistors) was provided in \mbox{ref.\ \cite{Bourjaily:2015jna}}; but a more general way to express it (independent of kinematical preferences) would be to start with a double-bubble---analogous to (\ref{bubble_definition}) above---and take two residues cutting the outermost amplitudes. 

This function represents the `correct result' for this two parameter function of the loop momenta, and so we must match field theory everywhere as a function of $(x,y)$. This can be done by simply matching field theory at an arbitrary (but fixed) point $(x^*,y^*)$. (These points can always be chosen so that the now-normalized basis integrand is dual-conformally-invariant, but dual-conformal-invariance is not something required by our approach.) Thus, we can match field theory at least at some arbitrarily chosen point $(x^*,y^*)$ along this co-dimension six residue of the integrand using the terms:
\vspace{-2.5pt}\eq{\mathcal{A}_n=\bigger{\sum}\left(\fwbox{60pt}{\fig{-24.6pt}{1}{two_loop_ladder_1}}\bigger{\times}\fwbox{65pt}{\fig{-24.6pt}{1}{two_loop_cut_1}}\hspace{3pt}\right)+\ldots\,,\label{first_partial_two_loop_expansion}\vspace{-2.5pt}}
To be completely clear, the scalar double-box integrands have been normalized (fixing their one degree of freedom) at some particular point $(x,y)$ to match the corresponding ($(x,y)$-dependent) point in field theory. We could include these labels in the figure denoting the integrand in our basis, but have left them off for notational simplicity. Also, the labels $x,y$ in the on-shell function should really be understood as $x^*,y^*$---where these particular points are fixed, but chosen arbitrarily.

The attentive reader will notice that these cuts, (\ref{two_loop_cut_1}), would also have support from the other integrals in our basis (\ref{two_loop_ints}). And so it would seem that we are in danger of spoiling the correctness of the terms (\ref{first_partial_two_loop_expansion}) on the cuts (\ref{two_loop_cut_1}) once we include the other integrals in our basis. This potential problem is easily solved by using the contact term degrees of freedom of the higher integrands in our basis to ensure that all the other integrals {\it vanish on these cuts}. Just to be clear, it is not possible to make these higher integrands vanish everywhere on the lower cuts, but only at the specified points, $(x^*,y^*)$ etc., along the lower cuts. Let us now describe how this works in detail. 

Consider now the pentaboxes in our basis of integrands (\ref{two_loop_ints}). Each of these has exactly two non-contact, and four contact degrees of freedom. Can these integrals be used to match field theory cuts not matched already by the terms in (\ref{first_partial_two_loop_expansion})? And can we do so without spoiling those already matched, (\ref{two_loop_cut_1})? The answer to both questions is yes. Let us consider each in turn. 

Pentabox integrals have more external propagators than the double-boxes, and therefore support field theory cuts involving more external propagators. We could use co-dimension seven cuts to match field theory in a way similar to what we did above, but (unlike the double-boxes), pentabox integrals {\it always} support `leading singularities'---residues of maximal co-dimension (eight)---for the simple reason that they involve eight total propagators. Whenever leading singularities are supported, they are better cuts to use---if only because they do not require any arbitrary choice of points such as $(x^*,y^*)$ on which to evaluate cuts of integrands and their on-shell function coefficients. Thus, the pentaboxes can be used to match (some of the) leading singularities of field theory:
\vspace{-15pt}\eq{\fwbox{85pt}{\fig{-24.6pt}{1}{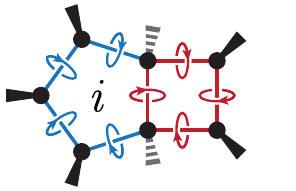}}\hspace{-10pt}\bigger{\leftrightarrow}\hspace{5pt}\fwbox{100pt}{\fig{-34.76pt}{1}{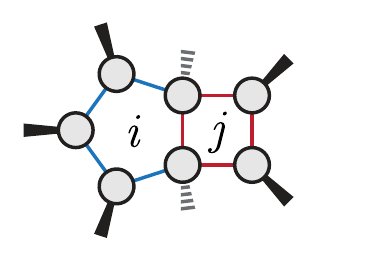}}\label{pentabox_cuts}\vspace{-10pt}}
When trying to match field theory on these cuts, there may seem to be a problem. On the one hand, the equations for cutting eight propagators has {\it four} distinct solutions (two per loop, labeled by $(i,j)$), and field theory residues evaluated at these points in loop-momentum space are generally distinct. On the other hand, there are only {\it two} non-contact degrees of freedom for the numerator of the pentabox. At best, we can match two of the four residues of field theory. (All $4$ of the contact term degrees of freedom vanish on these cuts by virtue of the fact that all of the contact terms involve fewer external propagators.)

The resolution of this problem is in fact simple: it is simply unnecessary to {\it manifestly} match {\it every} field theory residue. So long as we have a complete basis of integrands, and the coefficient of every integral in the basis is uniquely fixed by {\it some} residue, completeness of the basis ensures that all other residues will also be matched. Thus, we merely need to choose two of the four pentabox leading singularities to match manifestly using the $2$ non-contact-term degrees of freedom. Let us therefore declare that we fix the (non-contact-term) degrees of freedom of the pentaboxes in order to precisely match field theory on the `$j\!=\!1$' residues of field theory in (\ref{pentabox_cuts}):
\vspace{-10pt}\eq{\fwbox{100pt}{\fig{-34.76pt}{1}{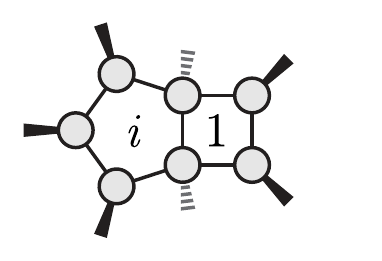}}\vspace{-10pt}\label{choice_of_pentabox_cuts}}

The above discussion has allowed us to uniquely specify the {\it non-contact-term} degrees of freedom of every pentabox integral, matching field theory on the (subset of) pentabox leading singularities in (\ref{choice_of_pentabox_cuts}). This can be done irrespective of the contact term degrees of freedom, as none of these terms have support on the residues (\ref{choice_of_pentabox_cuts}). Thus, we have a four-dimensional space of `ambiguities' for the possible numerators of the pentaboxes which leave in tact the correctness of the pentabox residues. The attentive reader should already understand how these degrees of freedom should be eliminated: following our general comments on prescriptive unitarity, these contact term degrees of freedom are eliminated in such a way that we {must} ensure that the pentabox integrals {vanish} on the already-matched points in loop-momentum space. Because each contact term of the pentabox corresponds to a double-box integral, and each of these have been used to match field theory at arbitrarily chosen points $(x,y)$, we now require that the pentabox integrals vanish at these points. 

These are homogeneous equations which are easy to solve analytically: one merely evaluates the non-contact-term numerators on the residues being used to define the contact-term integrals, and subtract. (Without this subtraction, our basis would be essentially upper-triangular in form, and so this subtraction represents the only `linear algebra' involved in our construction.)  For example, if the four external propagators of the pentabox are labeled $a,\ldots,d$, so that we may expand the numerator into the ``chiral'' basis given in (\ref{chiral_pentagon_basis}), then we simply define the total pentabox integral's numerator to be given by:
\vspace{-2.5pt}\eq{N^i(\ell)\equiv{\color{hblue}\x{\ell}{Y^i}}-\sum_{\fwbox{10pt}{\lambda\!\in\!\{a,b,c,d\}}}\frac{{\color{hblue}\x{\ell(x^*)}{{\color{hblue}Y^i}}}{\color{hred}\x{\ell}{\lambda}}}{\x{\ell(x^*)}{\lambda}}\,.\label{fully_specified_pentabox_numerator}\vspace{-2.5pt}}
Here, ${\color{hblue}\x{\ell}{{\color{hblue}Y^i}}}$ is one of the non-contact-term (``chiral'') numerators (generally ${\color{hblue}Y^i}$ is one of ${\color{hblue}Q^i_{abcd}}$ as in (\ref{chiral_pentagon_basis}), but normalized to match a particular cut in (\ref{choice_of_pentabox_cuts})); and $x^*$ is whatever point is used to define the double-box integrals in the basis---which need not be the same for every double-box. This is analogous to Gram-Schmidt orthogonalization. We are only being somewhat schematic here because any more concreteness would require some reference to formulae for ${\color{hblue}\x{\ell}{{\color{hblue}Y^i}}}$ (requiring in turn the need to introduce notation using some kinematic scheme---about which we desire to remain agnostic), and also a specific rule for specifying the points $(x^*,y^*)$ which are truly arbitrary. For a more concrete discussion, we refer the reader to \mbox{ref.\ \cite{Bourjaily:2015jna}}. 

Thus, we have now fully specified all pentabox integrands in our basis, and the coefficient of each is uniquely fixed by a single field theory residue. 

All that remains for us to do is choose which double-pentagons to include in our basis. As before, this can be done rather simply. Each double-pentagon has $4$ non-contact degrees of freedom, and---conveniently this time---has precisely four leading singularities not shared by any other integrals in our basis: the so-called `kissing boxes'. Thus, we may uniquely determine the non-contact degrees of freedom of the double-pentagons by ensuring that they match field theory on the four cuts: 
\vspace{-10pt}\eq{\fwbox{100pt}{\fig{-24.6pt}{1}{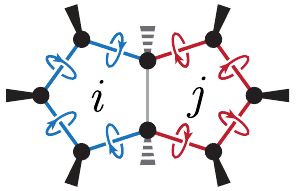}}\hspace{0pt}\bigger{\leftrightarrow}\hspace{5pt}\fwbox{100pt}{\fig{-34.76pt}{1}{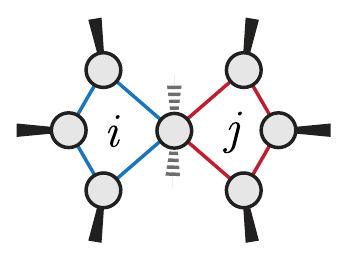}}\label{kissing_box_cuts}\vspace{-10pt}}
As before, the contact-terms for these integrals are fully determined by the requirement that these integrals vanish on all the cuts already matched by lower integrals. It is easy to see that exactly the right number of contact-term topologies exist to eliminate all these degrees of freedom. 

Having started with an over-complete basis of integrals, and defined each integrand uniquely to match field theory on a specific cut and to vanish on all other cuts used to define other integrals, we have achieved a truly diagonal basis. That this basis is not over-complete follows from the fact that each integrand has a unique field theory coefficient (because every other integral in the basis has been explicitly constructed to vanish there). Thus, we have found prescriptive representation of all two loop amplitude integrands in planar SYM. Schematically, we may write,
\vspace{-5pt}\eq{\mathcal{A}_n^{L=2}=\bigger{\displaystyle\sum_{\raisebox{2pt}{\scalebox{0.6}{$\!\mathcal{L}$}}}}f_{\mathcal{L}}\fwbox{90pt}{\fig{-24.62pt}{1}{two_loop_ladder_general}}\vspace{-5pt}\label{two_loop_amplitudes}}
where the `ladder' integrands are chosen from our basis (\ref{two_loop_ints}), constructed in the way described above, and each coefficient $f_{\mathcal{L}}$ is a a single on-shell function:
\vspace{-5pt}\eq{\begin{array}{@{}c@{}}\fwbox{90pt}{\fig{-24.62pt}{1}{two_loop_ladder_general}}\in\left\{\fwbox{70pt}{\fig{-24.6pt}{1}{two_loop_ladder_1}},\fwbox{80pt}{\hspace{10pt}\fig{-24.6pt}{1}{two_loop_ladder_2}},\fwbox{90pt}{\fig{-24.6pt}{1}{two_loop_ladder_3}}\right\}\\
\fwboxR{90pt}{f_{\mathcal{L}}}\in\left\{\fwbox{70pt}{\fig{-24.6pt}{1}{two_loop_cut_1}},\fwbox{80pt}{\hspace{14.9pt}\fig{-34.76pt}{1}{two_loop_cut_2}},\fwbox{90pt}{\fig{-34.76pt}{1}{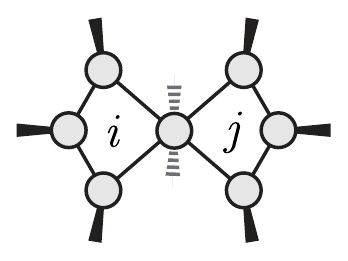}}\right\}\end{array}\vspace{-5pt}}

Readers familiar with the earlier work in \mbox{ref.\ \cite{Bourjaily:2015jna}} will notice that the representation of two loop amplitudes described here is considerably more compact (and arguably more straightforward). There are several reasons for this.

The primary distinction between the representation of two loop amplitudes in (\ref{two_loop_amplitudes}) and that described in \mbox{ref.\ \cite{Bourjaily:2015jna}} is that here we have made no use of composite residues such as (\ref{one_loop_composites}). Because these residues are entirely responsible for the infrared divergences of scattering amplitudes which are known to exponentiate according to the BDS ansatz described in \mbox{ref.\ \cite{Bern:2005iz}}, it is well-motivated to make this exponentiation manifest in an integrand-level representation. This was achieved in the formulation described in \mbox{ref.\ \cite{Bourjaily:2015jna}}, but at the cost of distinguishing the terms in (\ref{two_loop_amplitudes}) according to the possible masslessness of the external leg ranges of the integrals, and using different cuts/coefficients for the different cases---namely, using composite residues for the massless cases, and those similar to what we described above whenever composite residues would not exist. 

Our choice here to {\it not} make such distinctions is primarily pedagogical: breaking the basis of integrals into more cases requires a degree of unessential complication and a longer discussion. At three loops, choosing not to exploit composite residues leads to a considerably more compact formulation, but it is worth mentioning that we have been unable to make the exponentiation of infrared divergences manifest at three loops even if these distinctions had been made. As such, it is not merely the interest of brevity that motivates our choice to ignore any possible composite residues that may exist. Refining our representation of three loops to make the exponentiation of infrared divergences manifest would be an interesting exercise, but must be left for future research.

\newpage
\vspace{-0pt}\subsection{Generalities of a Prescriptive Approach to Unitarity}\label{subsec:generalities_of_prescriptive_unitarity}\vspace{-2pt}

We hope that the discussions above at one and two loops were sufficiently clear to illustrate the prescriptive approach to unitarity. In this section, we outline how this can be formulated for amplitudes in a truly general quantum field theory---without reference to planarity, supersymmetry, spacetime dimension, or even the masslessness of particles. We return to the case of one loop prescriptive unitarity in \mbox{section \ref{sec:one_loop_revisited}} in order to better illustrate how these methods work for theories with worse ultraviolet behavior than SYM. 

The first step is to construct a complete basis of (local) loop integrands, with numerators dictated by the power counting of the field theory in question. After an analogue of Passarino-Veltman reduction \cite{Passarino:1978jh}, an over-complete basis of integrands may be identified. From such a basis of integrands, a prescriptive representation for any amplitude would be found by the following procedure:  
\begin{enumerate}
\item draw all integrand topologies, dividing every numerator's degrees of freedom into non-contact terms and contact terms;
\item for each integrand, starting with those involving the fewest external propagators, specify an independent subset of field theory residues involving all external propagators, and use these to define its non-contact term degrees of freedom;  
\item fix each integrand's contact terms by the requirement that the integral vanish on all the residues used to define integrals with fewer propagators. 
\end{enumerate}

This procedure may be followed regardless of the power counting of the theory, its spacetime dimension, etc. The only annoyance that may arise is that the size of the basis may grow rapidly---requiring a correspondingly large number of cuts (some which may have identical topologies, but evaluated multiple points along their internal degrees of freedom). 

If the last step in this procedure above were ignored---so that the cuts which define each integral did not exactly correspond to a single field theory residue---then the actual coefficients could be easily found by linear algebra. In this case, what we have described would asymptotically amount to what was described by OPP in \mbox{ref.\ \cite{Ossola:2006us}} (where the coefficients of integrals represent the difference between the right answer and all coefficients of the higher-level integrals which pollute each cut in question). This distinction is perhaps better illustrated by example, and we refer the reader to a more thorough description of prescriptive unitarity at one loop in \mbox{section \ref{sec:one_loop_revisited}}. 

In order to find a truly prescriptive representation, it is the last step that is the most important. And it may not even be possible to satisfy. If care is not taken in the selection of cuts used to define the lower integrals, the requirement that higher integrals vanish on all cuts below may not be possible. This will happen whenever the cuts being used to define `lower' integrals outnumber the contact term degrees of freedom of integrals above. The easiest way to illustrate this potential problem is through concrete examples that first arise at three loops. Because of this, we refer the reader to \mbox{section \ref{subsubsec:very_carefully_chosen_cuts}} for a more thorough discussion.

Even without seeing the details of what can go wrong, we should emphasize that this potential tension is a very real one: no matter how cleverly cuts are chosen, it is not possible to avoid na\"{i}vely over-constraining the contact terms of some integrals at three loops. In the representation we describe in the next section, this seemingly over-constrained problem is in fact solvable, but such a solution was not guaranteed. As such, our result at three loops represents a non-trivial test that the prescriptive representations exist. 

What would it mean for the prescriptive approach {\it not} to work? This would happen if it were not possible to satisfy the requirement that some integrals' contact terms vanish on the all (supported) cuts used to define lower integrals. This would not be a fundamental problem, {\it per se}, as the integrand basis being generated would still be complete; and as such, there would surely exist a solution to generalized unitarity resulting in some representation for amplitudes of the form (\ref{schematic_unitarity_expansion}). The problem is that the representation that results would not be {\it prescriptive}. Why? Because, if higher integrals could not be made to vanish on some cut purportedly being used to define a lower integral, then this cut would have contributions from more than one integral in the basis. The basis would not be {\it diagonal} in cuts. As such, the coefficient of the lower integral would need to be the difference between the ``right answer in field theory'' and the sum of all the coefficients of higher integrals that pollute this cut. If there were only one such complication, this would not substantially complicate matters; if there were many, then the problem would revert to more complicated linear algebra---ubiquitous in a non-prescriptive approach. 

Because of this tension, it would be very interesting to see if prescriptive representations exist more generally---at higher orders of perturbation, for non-planar theories, for theories with less supersymmetry, etc. Even if prescriptive representations were not possible, however, we expect that the prescriptive {\it approach} described here would lead to a substantial improvement in the linear algebra involved in finding integrand-level representations of amplitudes.

\newpage
\vspace{-6pt}\section[Prescriptive Form for All Three Loop Amplitudes in Planar SYM]{Prescriptive Representation of All Three Loop Amplitudes}\label{sec:three_loop_result}\vspace{-10pt}
Let us now apply the prescriptive approach to construct a closed-form representation of all $n$-point N$^k$MHV amplitudes in planar SYM at three loops. Until now, the only cases known (for arbitrary multiplicity) were the three loop MHV integrands found in \cite{ArkaniHamed:2010kv,Bourjaily:2015jna}. In this section, we construct representations valid for all amplitudes,
\vspace{-13pt}\eq{\mathcal{A}_n^{L=3}=\bigger{\displaystyle\sum_{\raisebox{2pt}{\scalebox{0.6}{$\!\mathcal{W}$}}}}f_{\mathcal{W}}\,\,\fwbox{100pt}{\fig{-47.3pt}{1}{wheel_int_general}}\hspace{-10pt}\bigger{+}\hspace{5pt}\bigger{\displaystyle\sum_{\raisebox{2pt}{\scalebox{0.6}{$\!\mathcal{L}$}}}}f_{\mathcal{L}}\fwbox{130pt}{\fig{-34.76pt}{1}{ladder_int_general}}\,,\label{general_three_loop_rep_body}\vspace{-13pt}}
where the integrals span all contact-term topologies of those drawn above, and the non-contact terms of each are fully determined to match specific field theory cuts $f_{\mathcal{W}}$ and $f_{\mathcal{L}}$. As indicated in (\ref{general_three_loop_rep_body}), the possible integrands come in two principle topologies which we will call `wheels' and `ladders', respectively. In the next subsection, we will demonstrate that this corresponds to a complete basis for three loop integrands, and we give a complete enumeration of the contact-term topologies (relative to what is drawn in (\ref{general_three_loop_rep_body})) that appear in our basis. In the following subsection we illustrate the cuts which define our basis (and the field theory coefficients); complete details are provided in \mbox{Appendix \ref{appendix:details_three_loop_result}}. General aspects of (\ref{general_three_loop_rep_body}) are discussed in \mbox{section \ref{subsec:discussion_of_result}}.

\vspace{-9pt}\subsection{Constructing a Diagonal Integrand Basis for Three Loop Integrals}\label{subsec:three_loop_basis_summary}\vspace{-6pt}

As outlined in \mbox{section \ref{subsec:generalities_of_prescriptive_unitarity}}, the first step in applying prescriptive unitarity is to construct a complete basis of integrals (relevant to a particular quantum field theory). At two loops, we saw that all integrals (with the correct power counting for SYM) could be expanded into those involving at most four external propagators---generally, double-pentagon integrals and contact terms thereof. 

At three loops, the same rule applies: any integral involving more than four external propagators is expandable into those with fewer. For (single-loop-momentum) factors of integrands involving a single internal propagator, the argument is the same at two loops. New at three loops is the possibility that one loop involves two internal propagators. The fact that a heptagon involving five external and two internal propagators (with numerators spanning a 50-dimensional space according to (\ref{dimension_counting_by_power})) can be decomposed into those involving at most four external propagators is similarly obvious (in terms of counting), and easy to verify by counting. See \mbox{Table \ref{power_counting_table}} for more general counting. This fact demonstrates that general integrands with the wheel topology (the first terms in (\ref{general_three_loop_rep_body})) can involve at most four external propagators per loop, and that the ladder integrals drawn in (\ref{general_three_loop_rep_body}) are actually reducible into:
\vspace{10.pt}\eq{\begin{array}{@{}c@{}}~\\[-34pt]\raisebox{-0pt}{$\fwbox{120pt}{\fig{-34.76pt}{1}{ladder_int_general}}\bigger{\subset}\hspace{-1pt}\left\{\fwbox{120pt}{\hspace{-5pt}\fig{-34.76pt}{1}{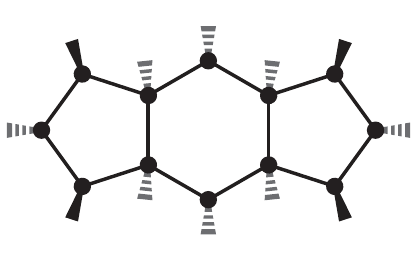}},\fwbox{100pt}{\fig{-34.76pt}{1}{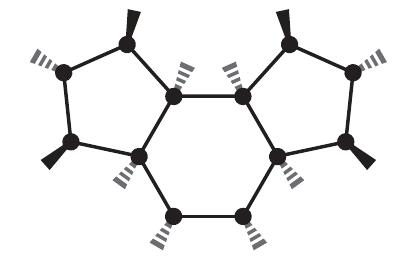}}\right\}\!.\label{decomposition_of_575_ladder}$}\\[-14pt]\end{array}\vspace{-20pt}}

One may wonder why we have excluded `wedge-type' integrals of the ladder topology---those involving four external propagators on one side of the middle loop. We do not list these because, without any loss of generality, they can always be considered as wheel integrals (by multiplying and dividing by the additional propagator). Thus, 
the integrals appearing in (\ref{decomposition_of_575_ladder}) together with the wheel in (\ref{general_three_loop_rep_body}) represents a (considerably over-)complete basis of integrands at three loops.

The second step in the procedure is to divide all the numerators for integrands in our basis into contact/non-contact-term degrees of freedom. In this partitioning, the only new cases to consider (relative to two loops) are the hexagons and pentagons involving two internal propagators. Let us describe the pentagons first. As should be familiar, the power counting of SYM dictates that these integrals involve numerators constructible as single inverse propagators. The division of this basis into contact/non-contact terms is obvious: if the three external propagators are labeled $a,b,c$, then we should use a basis of the form:
\vspace{-5pt}\eq{\x{\ell}{Y}\equiv(\ell-p_Y)^2\in\text{span}\underset{\text{``3+2'' basis for inverse propagators---three external}}{\big\{\overset{\text{{\color{hblue}non-contact}}}{{\color{hblue}\x{\ell}{Z^1}},{\color{hblue}\x{\ell}{Z^2}},{\color{hblue}\x{\ell}{Z^3}}},\overset{\text{{\color{hred}contact}}}{{\color{hred}\x{\ell}{a}},{\color{hred}\x{\ell}{b}},{\color{hred}\x{\ell}{c}}}\big\}}\,.\label{3_2_pentagon_basis}\vspace{-5pt}}
Here, the non-contact terms are somewhat schematic---they correspond to arbitrary inverse propagators ${\color{hblue}\x{\ell}{Z^I}}$ which span the three-dimensional space orthogonal to the three contact terms. The form of these numerators is not important, but the counting is. Thus, a pentagon integral involving three external and two internal propagators has ${\color{hblue}3}\!+\!{\color{hred}3}$ degrees of freedom---counting non-contact and contact terms, respectively. When indicating the three non-contact term degrees of freedom, we will use capital Roman letters $I,J,K\!\in\!\{1,2,3\}$ (letters corresponding to the different loops).

The final novelty to be discussed at three loops is the possibility of a hexagon involving four external and two internal propagators. This case turns out to be considerably simpler. Again, the power counting of SYM dictates that these integrals must involve 20 degrees of freedom, constructed as two-fold products of inverse propagators. A natural basis for these numerators is as follows:
\vspace{-8pt}\eq{\x{\ell}{Y}\x{\ell}{W}\in\text{span}\raisebox{5pt}{$\underset{\text{``chiral hexagon'' basis for two inverse propagators---four external}}{\raisebox{-5pt}{$\left\{\raisebox{27pt}{$\!\!$}\right.$}\begin{array}{l}\overset{\text{{\color{hblue}non-contact}}}{{\color{hblue}\x{\ell}{Q_{abcd}^1}^2},{\color{hblue}\x{\ell}{Q_{abcd}^2}^2}},\overset{\text{{\color{hred}contact}}}{{\color{hred}\x{\ell}{a}\x{\ell}{Z_{bcd}^I}},{\color{hred}\x{\ell}{b}\x{\ell}{Z^I_{acd}}}},
\\{\color{hred}\x{\ell}{c}\x{\ell}{Z^I_{abd}}},{\color{hred}\x{\ell}{d}\x{\ell}{Z^I_{abc}}},{\color{hred}\x{\ell}{a}\x{\ell}{b}},{\color{hred}\x{\ell}{a}\x{\ell}{c}},\\
{\color{hred}\x{\ell}{a}\x{\ell}{d}},{\color{hred}\x{\ell}{b}\x{\ell}{c}},{\color{hred}\x{\ell}{b}\x{\ell}{d}},{\color{hred}\x{\ell}{c}\x{\ell}{d}}\end{array}\raisebox{-5pt}{$\left.\raisebox{27pt}{$\!\!$}\right\}$}}$}.\label{4_2_hexagon_basis}\vspace{-6pt}}
Thus, hexagons have ${\color{hblue}2}$ non-contact- and ${\color{hred}18}$ contact-term degrees of freedom. Because they have only 2 non-contact terms, we will distinguish them by lower-case Roman letters $i,j,k\!\in\!\{1,2\}$ (again, the letters used to distinguish the loop momenta). 

We are now ready to enumerate all the possible topologies required in our basis, and count the non-contact term degrees of freedom of each. This is given in \mbox{Table \ref{table_of_integrals}}. The cuts used to define these integrals (and hence their field theory coefficients) are described in \mbox{Appendix \ref{appendix:details_three_loop_result}}.

\newpage
\begin{table}[h]
\vspace{-10pt}\eq{\fwbox{0pt}{\begin{array}{|@{$\;$}l@{$\hspace{-22.5pt}$}c@{$\hspace{-12.5pt}$}|@{$\;$}l@{$\hspace{-25pt}$}c@{$\hspace{-10pt}$}|}
\multicolumn{4}{c}{\text{`wheel' integrals $\mathcal{W}_i$}\fwbox{0pt}{\phantom{\mathcal{L}_i}}}\\
\hline\raisebox{30pt}{$\mathcal{W}_1\fwbox{0pt}{\phantom{\mathcal{L}_i}}$}&\fwbox{110pt}{\fig{-34.76pt}{1}{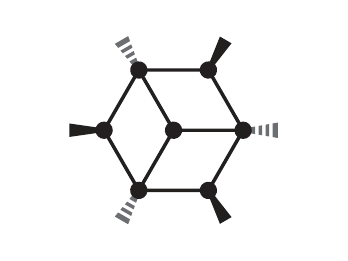}}&\raisebox{30pt}{$\mathcal{W}_2$}&\fwbox{110pt}{\fig{-34.76pt}{1}{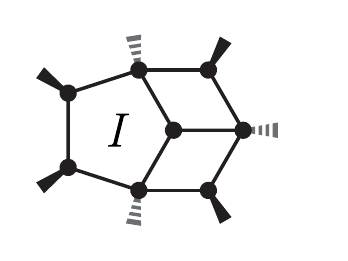}}\\
\hline\raisebox{30pt}{$\mathcal{W}_3$}&\fwbox{110pt}{\fig{-34.76pt}{1}{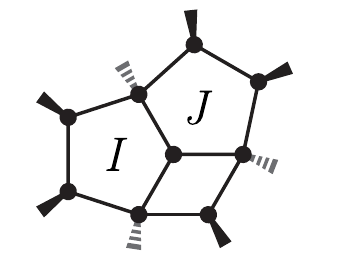}}&\raisebox{30pt}{$\mathcal{W}_4$}&\fwbox{110pt}{\hspace{10pt}\fig{-34.76pt}{1}{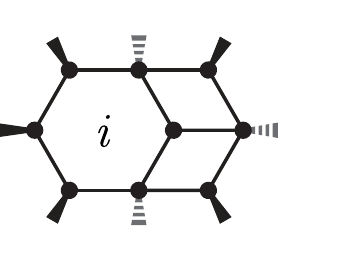}}\\
\hline\raisebox{40pt}{$\mathcal{W}_5$}&\fwbox{110pt}{\fig{-47.3pt}{1}{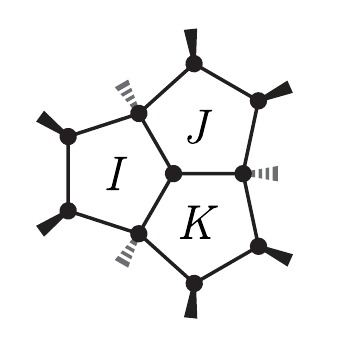}}&\raisebox{40pt}{$\mathcal{W}_6$}&\fwbox{110pt}{\fig{-47.3pt}{1}{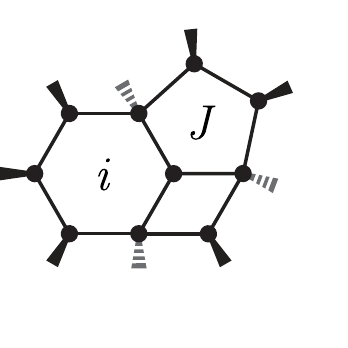}}\\
\hline\raisebox{40pt}{$\mathcal{W}_7$}&\fwbox{110pt}{\fig{-47.3pt}{1}{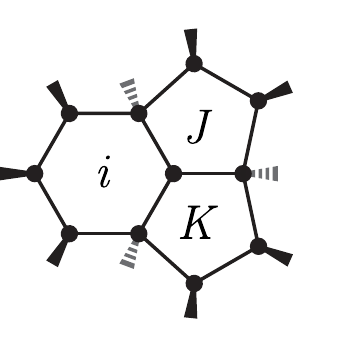}}&\raisebox{40pt}{$\mathcal{W}_8$}&\fwbox{110pt}{\fig{-47.3pt}{1}{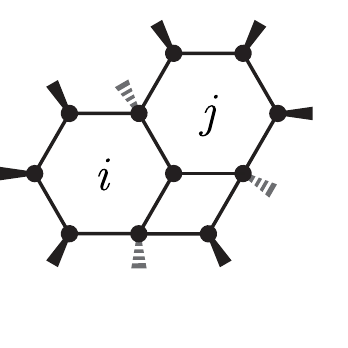}}\\
\hline\raisebox{40pt}{$\mathcal{W}_9$}&\fwbox{110pt}{\fig{-47.3pt}{1}{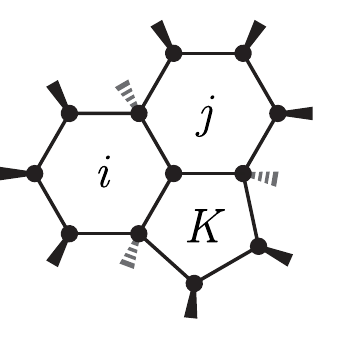}}&\raisebox{40pt}{$\mathcal{W}_{10}$}&\fwbox{110pt}{\fig{-47.3pt}{1}{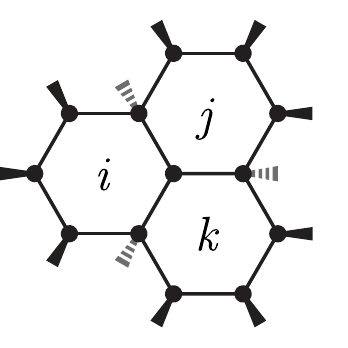}}\\
\hline
\end{array}
\hspace{1pt}\begin{array}{|@{$\;$}l@{$\hspace{-20.pt}$}c@{$\hspace{-0pt}$}|@{$\;$}l@{$\hspace{-25pt}$}c@{$\hspace{-5pt}$}|}
\multicolumn{4}{c}{\text{`ladder' integrals $\mathcal{L}_i$}\fwbox{0pt}{\phantom{\mathcal{W}_i}}}\\[0.05pt]
\hline\raisebox{30pt}{$\mathcal{L}_{1}\fwbox{0pt}{\phantom{\mathcal{W}_i}}$}&\fwbox{120pt}{\fig{-34.76pt}{1}{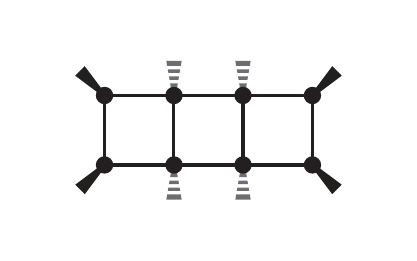}}&\raisebox{30pt}{$\mathcal{L}_{2}$}&\fwbox{120pt}{\fig{-34.76pt}{1}{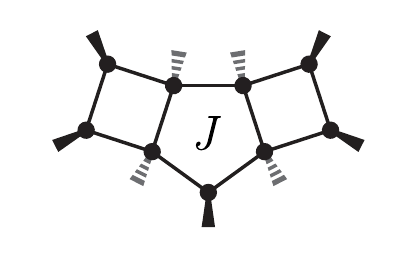}}\\
\hline\raisebox{30pt}{$\mathcal{L}_{3}$}&\fwbox{120pt}{\fig{-34.76pt}{1}{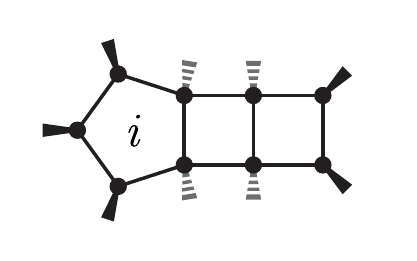}}&\raisebox{30pt}{$\mathcal{L}_{4}$}&\fwbox{120pt}{\fig{-34.76pt}{1}{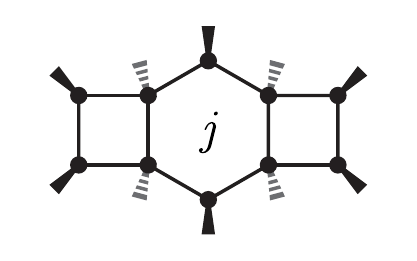}}\\
\hline\raisebox{30pt}{$\mathcal{L}_{5}$}&\fwbox{120pt}{\fig{-34.76pt}{1}{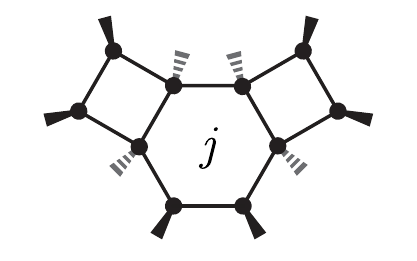}}&\raisebox{30pt}{$\mathcal{L}_{6}$}&\fwbox{120pt}{\fig{-34.76pt}{1}{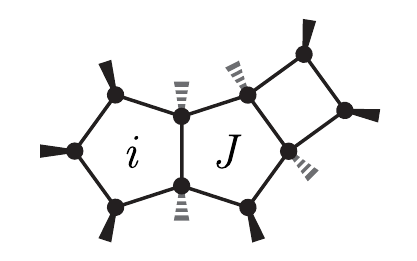}}\\
\hline\raisebox{30pt}{$\mathcal{L}_{7}$}&\fwbox{120pt}{\fig{-34.76pt}{1}{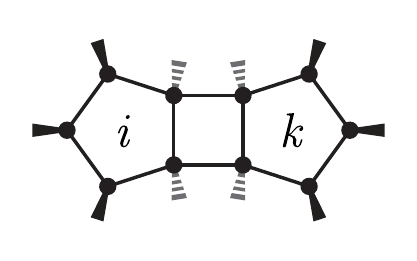}}&\raisebox{30pt}{$\mathcal{L}_{8}$}&\fwbox{120pt}{\fig{-34.76pt}{1}{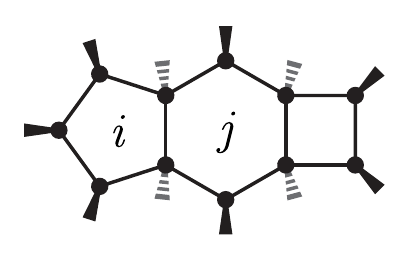}}\\
\hline\raisebox{30pt}{$\mathcal{L}_{9}$}&\fwbox{120pt}{\fig{-34.76pt}{1}{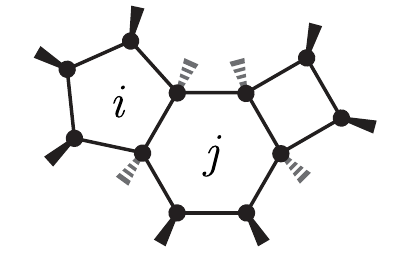}}&\raisebox{30pt}{$\mathcal{L}_{10}$}&\fwbox{120pt}{\fig{-34.76pt}{1}{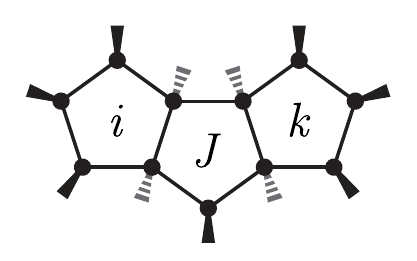}}\\
\hline\raisebox{30pt}{$\mathcal{L}_{11}$}&\fwbox{120pt}{\fig{-34.76pt}{1}{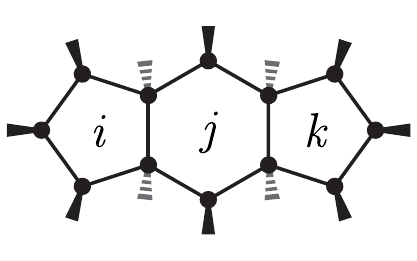}}&\raisebox{30pt}{$\mathcal{L}_{12}$}&\fwbox{120pt}{\fig{-34.76pt}{1}{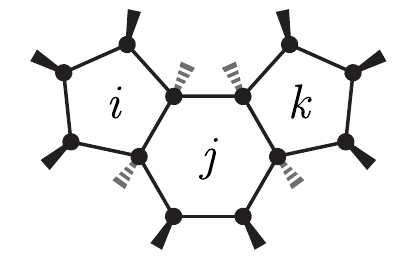}}\\[-0.45pt]
\hline
\end{array}}\nonumber\vspace{-12pt}}
\caption{The integral topologies which form a complete basis for three loop amplitudes in planar SYM. Here, $\{i,j,k\}\!\in\!\{1,2\}$ and $\{I,J,K\}\!\in\!\{1,2,3\}$ label non-contact degrees of freedom for general numerators, explicitly matching the on-shell functions in \mbox{Appendix \ref{appendix:details_three_loop_result}}.\label{table_of_integrals}}\vspace{-5pt}
\end{table}

For the sake of clarity, each integral topology drawn in \mbox{Table \ref{table_of_integrals}} represents the collection of all integrals with distinct, cyclically-ordered leg distributions. (Loop momentum labels are always symmetrized.) For most integrals, asymmetry in the diagram is compensated by rotational invariance in sum---for example, the reflected images of $\mathcal{L}_2$ or $\mathcal{L}_3$ are already accounted for. However, some reflected integrals should be considered implicit: namely, the reflected images of: $\mathcal{W}_6$, $\mathcal{L}_6$, and $\mathcal{L}_9$---for which the defining cuts related by symmetry to those drawn in \mbox{Appendix \ref{appendix:details_three_loop_result}}. 

\newpage
\vspace{-6pt}\subsection{Illustrations of Integrand-Defining Cuts and Coefficients}\label{subsec:exempli_cut_rules}\vspace{-6pt}
As mentioned above, the full list of cuts used to define each integral in our basis in \mbox{Table \ref{table_of_integrals}} together with the coefficient of each is described in \mbox{Appendix \ref{appendix:details_three_loop_result}}. In this section we merely illustrate some examples of these defining cuts and corresponding coefficients. We start with the most obvious and then discuss some truly arbitrary choices made before addressing some of the more subtle issues that are involved. 

These subtleties arise because some of the cuts necessarily or potentially used to define the wheel-topology integrals have support as contact terms of the ladder-topology integrals. We will see that some of this overlapping support is necessary and important, but also has the potential to spoil the diagonalizability of our basis. Indeed, we will see that for exactly one of the integrals in our basis, $\mathcal{W}_5$, this cross-talk between topologies poses a critical and unavoidable tension that, if unresolved, could spoil the existence of any prescriptive representation. For this reason, this integral's defining cuts will be described in some detail, making clear how this tension arises and how it is resolved. 

\vspace{-8pt}\subsubsection{Obvious or Arbitrary Choices for Cuts and Coefficients}\label{subsubsec:arbitrary_choices_for_cuts}\vspace{-6pt}
Analogous to the double-pentagon integrals at two loops, some of the integrals in our basis are defined by entirely obvious cuts. This is the case for the top-level integrals in our basis: $\mathcal{W}_{10}$, $\mathcal{L}_{11}$ and $\mathcal{L}_{12}$,
\vspace{-10pt}\eq{\left\{\raisebox{45pt}{$\!\!$}\right.\fwbox{100pt}{\fig{-47.3pt}{1}{wheel_int_10}}\hspace{-5pt},\fwboxL{120pt}{\fig{-34.76pt}{1}{ladder_int_11}},\fwboxL{110pt}{\hspace{-5pt}\fig{-34.76pt}{1}{ladder_int_12}}\left.\raisebox{45pt}{$\!\!$}\right\}\!.\label{top_level_ints_with_obvious_cuts}\vspace{-10pt}}
Each of these integrals have precisely twelve external propagators, giving rise to leading singularities indexed by $\{i,j,k\}\!\in\!\{1,2\}$ indicating the two solutions to each one-loop box:
\vspace{-10pt}\eq{\left\{\raisebox{45pt}{$\!\!$}\right.\fwbox{100pt}{\fig{-47.3pt}{1}{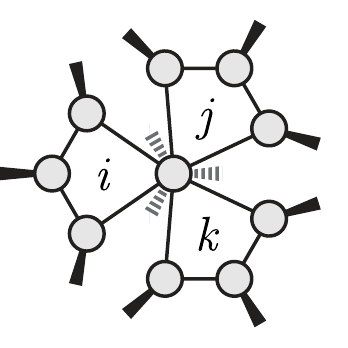}}\hspace{-5pt},\fwboxL{125pt}{\fig{-34.76pt}{1}{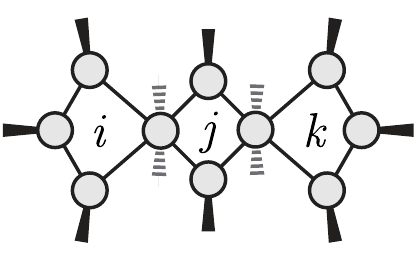}},\hspace{-5pt}\fwboxL{110pt}{\hspace{-5pt}\fig{-34.76pt}{1}{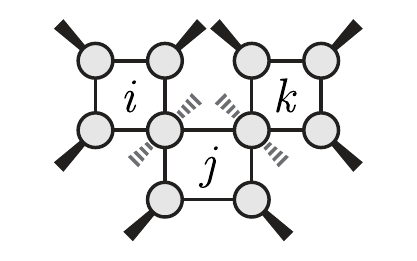}}\left.\raisebox{45pt}{$\!\!$}\right\}\!.\label{obvious_cuts_of_top_level_ints}\vspace{-10pt}}
In each case, the $2^3$ leading singularities can be used to define the corresponding integral's $2^3$ non-contact degrees of freedom in its numerator.

Let us now consider a case where some choices of cuts must be made, but where this choice is completely arbitrary. Among the simplest examples where such a choice is required happens for the wheel integral $\mathcal{W}_9$. In this case, there are only eleven external propagators, and so some internal propagator must also be cut to give a leading singularity. There are two potential topologies of these residues, depending on which internal propagator is cut:\footnote{The third propagator cannot be cut in a leading singularity as it would require more than four constraints to be imposed on a single loop momentum.}
\vspace{-10pt}\eq{\left\{\raisebox{45pt}{$\!\!$}\right.\fwbox{100pt}{\fig{-47.3pt}{1}{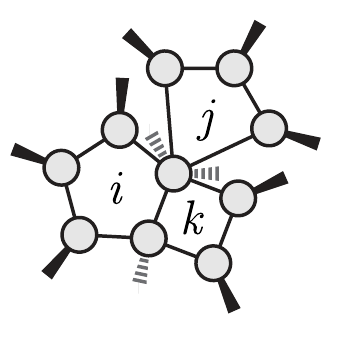}}\hspace{-5pt},\fwbox{100pt}{\hspace{5pt}\fig{-47.3pt}{1}{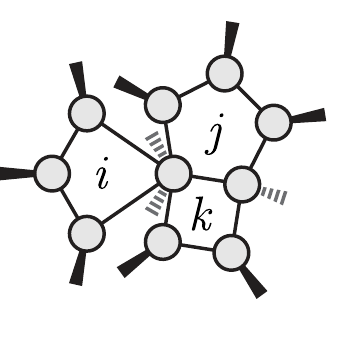}}\left.\raisebox{45pt}{$\!\!$}\right\}\!.\label{potential_w9_cuts}\vspace{-10pt}}
There are therefore $2\!\times\!2^3$ natural leading singularities to match, but only $12$ (non-contact) degrees of freedom in the numerator. Thus, it is simply not possible to construct a numerator for $\mathcal{W}_{9}$ for which each of the leading singularities (\ref{potential_w9_cuts}) are matched identically.

This situation is analogous to the case of the pentabox leading singularities and integrals at two loops (see \mbox{section \ref{subsubsec:diagonal_basis_of_cuts}}). And the solution is the same: it simply does not matter which choice of cuts is used to match field theory---the non-manifestly matched cut(s) will always follow from completeness of our basis. Thus, we have simply chosen to fix $k\!=\!1$ for the second topology in (\ref{potential_w9_cuts}), matching the 12 leading singularities,
\vspace{-10pt}\eq{\left\{\raisebox{45pt}{$\!\!$}\right.\fwbox{100pt}{\fig{-47.3pt}{1}{wheel_cut_9a}}\hspace{-5pt},\fwbox{100pt}{\hspace{5pt}\fig{-47.3pt}{1}{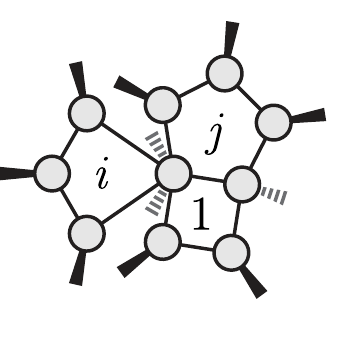}}\left.\raisebox{45pt}{$\!\!$}\right\}\!.\label{chosen_w9_cuts}\vspace{-10pt}}

It is worth seeing how the `missing' leading singularities are matched indirectly through a residue theorem. In this case, the residue theorem is:
\vspace{-10pt}\begin{align}&\hspace{12.5pt}\bigger{\partial}\left[\raisebox{45pt}{$$}\right.\hspace{-5pt}\fwbox{95pt}{\hspace{5pt}\fig{-47.3pt}{1}{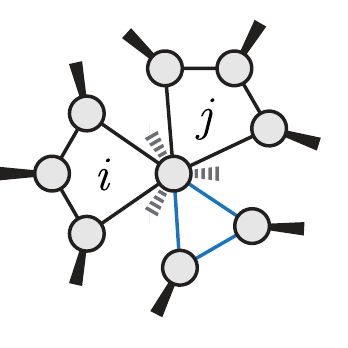}}\left.\raisebox{45pt}{$\!\!\!$}\right]=0\label{eg_grt}\\[-15pt]&\!=\!\!\sum_k\!\!\left\{\raisebox{45pt}{$\!\!$}\right.\hspace{-5pt}\fwbox{100pt}{\hspace{5pt}\fig{-47.3pt}{1}{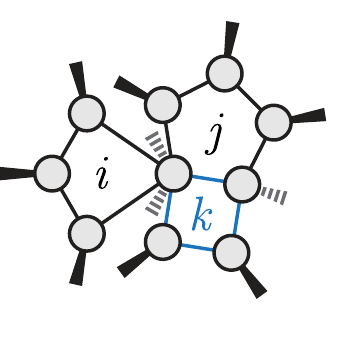}}\hspace{-14pt}+\hspace{-6pt}\fwbox{100pt}{\hspace{5pt}\fig{-47.3pt}{1}{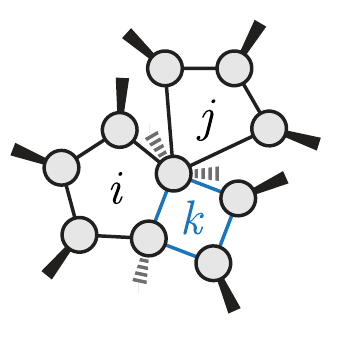}}\hspace{-10pt}+\fwbox{95pt}{\hspace{5pt}\fig{-47.3pt}{1}{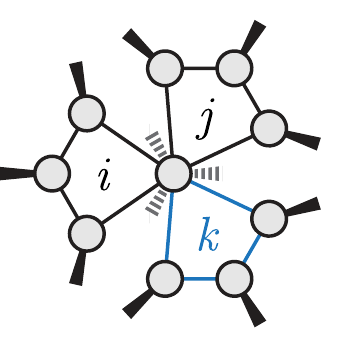}}\hspace{-5pt}+\fwbox{90pt}{\hspace{-2pt}\fig{-47.3pt}{1}{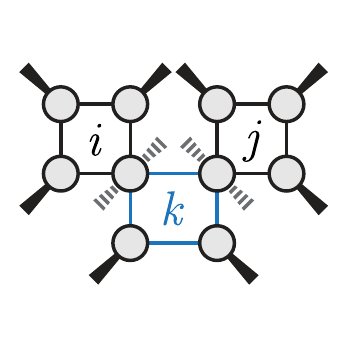}}\left.\raisebox{45pt}{$\!\!\!$}\right\}\nonumber\\[-30pt]\nonumber\end{align}
Fixing the solutions $(i,j)$ of the two quadruple cuts, we start from the one-parametric function depicted in the first line of (\ref{eg_grt}) and sum over all allowed factorization channels (including the different solutions labeled by $k$). This is simply a manifestation of Cauchy's theorem. Notice that the first term in the summand (\ref{eg_grt}) includes the both the `missing' $k\!=\!2$ cuts and the `matched' $k\!=\!1$ residues of our choice (\ref{chosen_w9_cuts}), and every other residue appearing in this theorem has been matched explicitly. Thus, this identity directly allows us to express the unmatched cut in terms of those we have matched. 

Of course, in order for this to work, every integrand supporting the other cuts must have support on the unmatched cut:
\vspace{-10pt}\eq{\fwbox{100pt}{\fig{-47.3pt}{1}{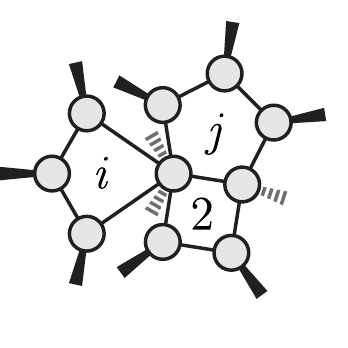}}\label{w9_unmatched_cut}\vspace{-10pt}}
Interestingly, once the non-contact degrees of freedom of $\mathcal{W}_9$ have been fixed according to the choice (\ref{chosen_w9_cuts}), every one of these integrals {\it automatically} contributes (with a minus sign) on the unmatched cut; and similarly, once the contact terms of $\mathcal{W}_{10}$ and $\mathcal{L}_{12}$ have been fixed by the requirement that these integrals vanish on the cuts in (\ref{chosen_w9_cuts}), these integrals automatically have support on the unmatched cuts (\ref{w9_unmatched_cut}) as well. Thus, every term required in the residue theorem (\ref{eg_grt}) does contribute support on the non-manifestly-matched cuts, with the requisite signs in order to exactly match field theory on the non-manifestly-matched residue (\ref{w9_unmatched_cut}).

Such residue theorems are fun to illustrate, but the fact that {\it some} residue theorem ensures that any non-manifestly-matched cut of field theory works follows automatically from completeness. Hence, we will spare the reader the (somewhat tedious) exercise of describing how this works in every particular case that follows.

\vspace{-0pt}\subsubsection{Somewhat Carefully Chosen Cuts and Coefficients}\label{subsubsec:somewhat_carefully_chosen_cuts}\vspace{-2pt}

No wheel integral has support on a cut used to define a ladder; but the converse is not true. In fact, we have already seen this in action: the cuts used to define integral $\mathcal{W}_9$ have support from $\mathcal{L}_{12}$; and the requirement that $\mathcal{L}_{12}$ vanish on these cuts precisely accounts for all its its wedge-type contact-term degrees of freedom. 

This happens frequently, but requires a minimal degree of care. This is perhaps best illustrated by example. Consider the wheel integral $\mathcal{W}_2$. It has 3 (non-contact) degrees of freedom that must be fixed. Cutting all propagators of the diagram results in $2\!\times\!2$ cuts:
\eq{\fwbox{85pt}{\fig{-34.76pt}{1}{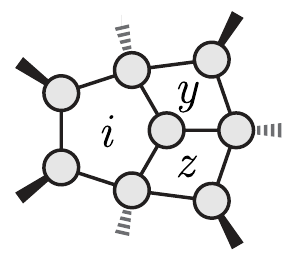}}\bigger{\subset}\left\{\fwbox{85pt}{\fig{-34.76pt}{1}{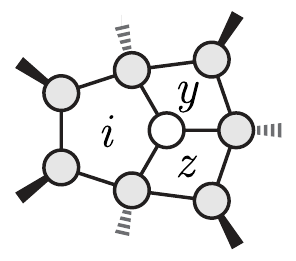}},\fwbox{85pt}{\fig{-34.76pt}{1}{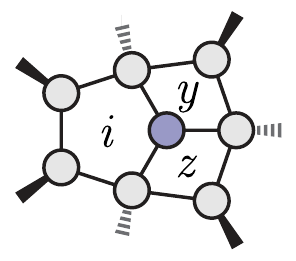}}\right\}\label{w2_maximal_cuts}}
Here, the blue and white three-point vertices represent MHV and $\overline{\text{MHV}}$ amplitudes, respectively. The choice we make in \mbox{Appendix \ref{appendix:details_three_loop_result}} is perhaps the obvious one: simply choose 3 of the 4 possible cuts in (\ref{w2_maximal_cuts}),
\vspace{-2pt}\eq{\left\{\fwbox{85pt}{\fig{-34.76pt}{1}{wheel_cut_2_general_a}},\fwbox{85pt}{\fig{-34.76pt}{1}{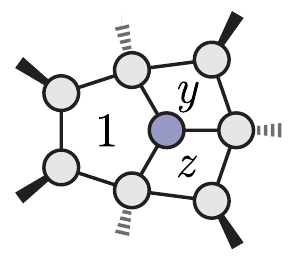}}\right\}\,.\label{chosen_w2_cuts}\vspace{-2pt}}

Although this choice works, it is worth illustrating an alternative choice that may appear acceptable but that would in fact have been problematic. As far as the non-contact degrees of freedom of $\mathcal{W}_2$ are concerned, any three independent cuts involving all external propagators would suffice. What would have been wrong with the following choice:
\vspace{-10pt}\eq{\fwbox{85pt}{\fig{-34.76pt}{1}{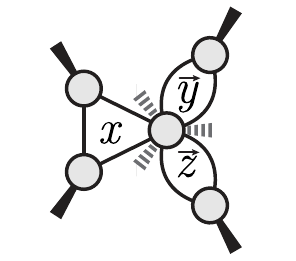}}\quad\text{instead of}\fwbox{85pt}{\fig{-34.76pt}{1}{wheel_cut_2_general_b1}}\,?\label{bad_w2_cuts}\vspace{-10pt}}
The cut on the left in (\ref{bad_w2_cuts}) does indeed determine the remaining non-contact degree of freedom of $\mathcal{W}_2$ as well as that on the right. As far as the wheel integrals are concerned, any wheel with support on one will have support on the other; and so, this choice has no effect on the constraints imposed for the contact terms of higher wheel integrals. The problem, however, is that some ladder integrals have support on the left-hand cut in (\ref{bad_w2_cuts}), but not on the right-hand choice. Moreover, it is easy to see that there do not exist contact-term degrees of freedom for ladder integrals capable of making these vanish on the left-hand cut in (\ref{bad_w2_cuts}). This means that first choice would spoil the diagonalizability of our basis. 

A similar situation arises for the wheel $\mathcal{W}_1$: it has a single degree of freedom, and it may have seemed convenient to fix this along a cut with the topology of three `kissing' bubbles:
\eq{\fwbox{85pt}{\fig{-34.76pt}{1}{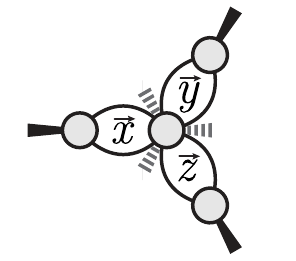}}\label{kissibg_bubbles}}
Choosing arbitrary points $(\vec{x},\vec{y},\vec{z})$ along this residue could indeed be used to define the single degree of freedom in the numerator of $\mathcal{W}_1$. However, this cut topology has support from {\it many} ladder integrals---which cannot be made to simultaneously vanish on these cuts; it would over-constrain the contact-term degrees of freedom of the ladders---for example $\mathcal{L}_2$. Thus, we cannot choose to define $\mathcal{W}_1$ by the cut (\ref{kissibg_bubbles}). In order to avoid over-constraining the contact-term degrees of freedom of the ladder integrals, it is necessary for us to ensure that the cut used to define $\mathcal{W}_{1}$ has no support on any of the ladders. This is only achieved if all the internal propagators of $\mathcal{W}_1$ are cut when defining its normalization and coefficient. 

Cutting every propagator of $\mathcal{W}_1$ results in 2 possible solutions (each parameterized by three internal degrees of freedom $(x,y,z)$ which must be chosen arbitrarily), distinguished by the MHV-degree of the internal, three-point amplitude. The choice between which of these two cuts should be used to define $\mathcal{W}_1$ and its coefficient is arbitrary, but must be made. We have chosen the former:
\vspace{-20pt}\eq{\fwbox{100pt}{\fig{-47.3pt}{1}{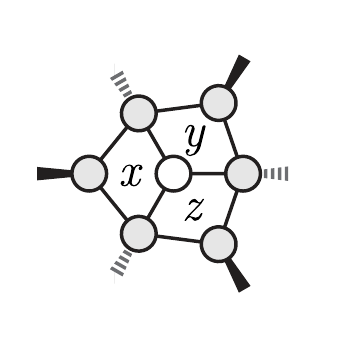}}\vspace{-20pt}\label{chosen_w1_cut}}

The general rule to avoid these potential problems should now be obvious: cut as many internal propagators as possible to define as many non-contact degrees of freedom of any integral---making sure that the number of cuts used with a given topology do not exceed the degrees of freedom of any potential contact terms from higher integrals (especially with different non-contact topologies). Thus, whenever a cut used to define a wheel integral that has support from (the contact terms of) ladder integrals, the number of cuts should not exceed the degrees of freedom of the overlapping contact terms. 

It is relatively easy to verify that the defining cuts of wheel integrals with support from ladder integral contact-terms given in \mbox{Appendix \ref{appendix:details_three_loop_result}} exactly accounts for the right counting, with exactly one unavoidable exception. This exception, the resolution of the resulting tension, and its potential implications for (the viability of) prescriptive unitarity more generally are discussed presently.

\vspace{-0pt}\subsubsection{Very Carefully Chosen Cuts: Magic Needed, Magic Found}\label{subsubsec:very_carefully_chosen_cuts}\vspace{-2pt}

As mentioned above, cutting as many internal propagators as possible to define the wheel integrals works quite well, with one important exception. While easy to overlook, its potential implications beyond three loops (and for more general theories) warrants a more thorough discussion. 

The exceptional case is for the wheel $\mathcal{W}_5$ consisting of three pentagon integrals:
\vspace{-10pt}\eq{\fwbox{100pt}{\fig{-47.3pt}{1}{wheel_int_5}}\label{wheel_int_5}\vspace{-10pt}}
This integral has $3^3\!=\!27$ non-contact degrees of freedom that we must fix by cuts. Following the rule described above, it is natural to start with the leading singularities---those cuts involving putting all 12 propagators on-shell:
\vspace{-10pt}\eq{\fwbox{90pt}{\fig{-47.3pt}{1}{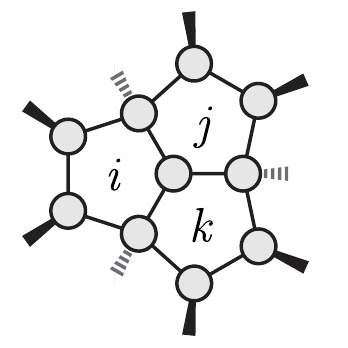}}\bigger{\subset}\left\{\raisebox{51pt}{$\!\!$}\right.\fwbox{90pt}{\fig{-47.3pt}{1}{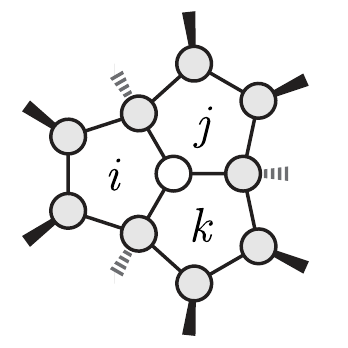}},\fwbox{90pt}{\fig{-47.3pt}{1}{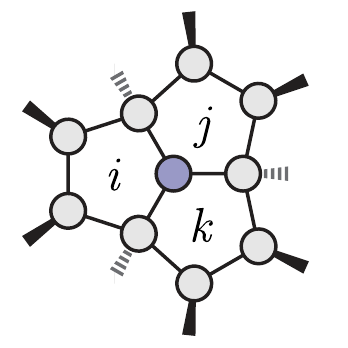}}\left.\raisebox{51pt}{$\!\!$}\right\}\,.\vspace{-10pt}\label{w5_cuts_a}}
It is easy to verify explicitly that of these 16 leading singularities, only 15 are independent points in the space of numerators. Thus, any choice of 15 can be used to define this number of non-contact degrees of freedom of the $\mathcal{W}_5$ numerators, leaving us with 12 degrees of freedom. We have chosen to match all cuts (\ref{w5_cuts_a}) except the following:
\vspace{-10pt}\eq{\fwbox{300pt}{\hspace{-30pt}\fwbox{90pt}{\fig{-47.3pt}{1}{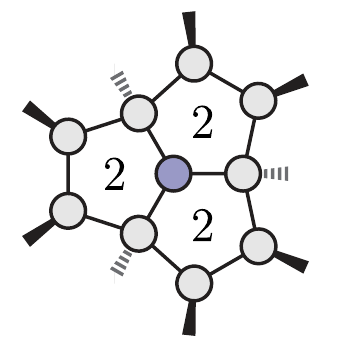}}\hspace{-5pt}\bigger{\neq}\fwbox{85pt}{\fig{-47.3pt}{1}{wheel_cut_5a}}\hspace{-5pt}\bigger{\subset}\left\{\raisebox{51pt}{$\!\!$}\right.\fwbox{85pt}{\fig{-47.3pt}{1}{wheel_cut_5a1}},\fwbox{90pt}{\fig{-47.3pt}{1}{wheel_cut_5a2}}\hspace{-5pt}\left.\raisebox{51pt}{$\!\!$}\right\}\!\!.}\vspace{-10pt}\label{choice_of_w5_cuts}}

This selection is truly arbitrary: any choice is equally valid, without causing complications. The subtlety (and true tension) arises in the choice of the cuts that define the remaining 12 degrees of freedom of $\mathcal{W}_5$. Because we have exhausted the leading singularities in (\ref{w5_cuts_a}), these additional cuts must leave some internal propagators uncut, and therefore have the topology of wedges. Arguably the most obvious choice for fixing the remaining degrees of freedom would be the following $4\!\times\!3$ cuts,
\vspace{-10pt}\eq{\left\{\raisebox{45pt}{$\!\!$}\right.\fwbox{90pt}{\fig{-47.3pt}{1}{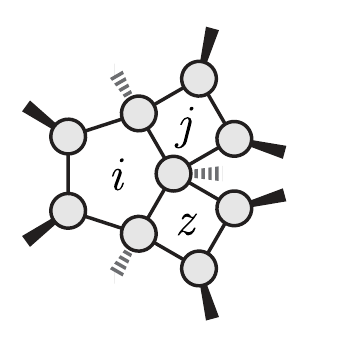}}\hspace{-10pt}\left.\raisebox{45pt}{$\!\!$}\right\}+ \text{cyclic.}\vspace{-10pt}\label{w5_cuts_b}}
These cuts are indeed independent and can be used to fully define the last 12 non-contact degrees of freedom of the $\mathcal{W}_5$ numerators. The problem is that we must ensure that all {\it other} integrals' contact terms vanish for the cuts being used to define the integrals in our basis. The relevant contact terms to consider in this case are from the ladders---for example, those of $\mathcal{L}_5$, which has 3 contact-term degrees of freedom with the topology of a wedge integral exactly involving the propagators in (\ref{w5_cuts_b}). 
\vspace{-7pt}\eq{\fwbox{105pt}{\fig{-34.76pt}{1}{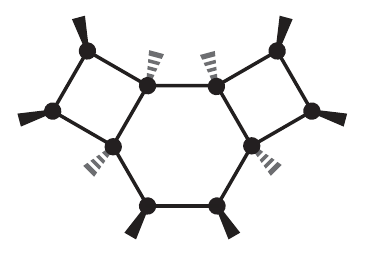}}\bigger{\supset}\fwbox{85pt}{\fig{-34.76pt}{1}{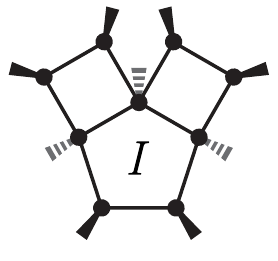}}\vspace{-7pt}\label{wedge_int_example}}
These contact terms have 3 degrees of freedom each, and cannot be made to vanish on all the 4 cuts of (\ref{w5_cuts_b}). 

This problem is in fact unavoidable, with no obvious solution. No matter what 12 cuts (besides the 15 in (\ref{w5_cuts_a})) are used to define the non-contact degrees of freedom of the $\mathcal{W}_5$ integrals, they will necessarily have 4 cuts supported by wedge-type contact terms of the ladders as in (\ref{wedge_int_example}). It is not hard to verify that the $4\!\times\!3$ cuts in (\ref{w5_cuts_b}) leads to an over-constrained problem without a solution; and breaking cyclicity will not help. Does there exist another choice of cuts for which a solution to the over-constrained system exists? 

The answer is yes: the choice presented in \mbox{Appendix \ref{appendix:details_three_loop_result}} does work. We do not wish to claim uniqueness of this solution to this potential obstruction, but it is worth mentioning that many other choices were tried (none of which worked). The resolution we found chooses only 3 of the 4 cuts in (\ref{w5_cuts_b}) for each cyclic image, 
\vspace{-10pt}\eq{\left\{\raisebox{45pt}{$\!\!$}\right.\fwbox{90pt}{\fig{-47.3pt}{1}{wheel_cut_5b}}\hspace{-7pt}\bigger{\neq}\fwbox{90pt}{\fig{-47.3pt}{1}{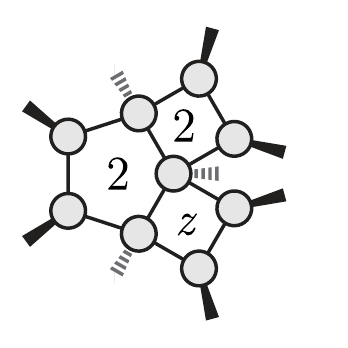}}\hspace{-10pt}\left.\raisebox{45pt}{$\!\!$}\right\}+ \text{cyclic,}\vspace{-10pt}\label{actual_w5_cuts_b}}
allowing us to fix $3\!\times\!3$ of the remaining 12 degrees of freedom of $\mathcal{W}_5$. The final 3 degrees of freedom are then fixed by lower cuts, 
\vspace{-18pt}\eq{\left\{\raisebox{40pt}{$\!\!$}\right.\fwbox{75pt}{\fig{-47.3pt}{1}{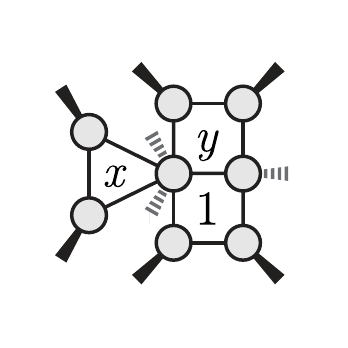}}\hspace{-2pt}\left.\raisebox{40pt}{$\!\!$}\right\}+ \text{cyclic.}\vspace{-18pt}\label{actual_w5_cuts_c}}
We should mention that this choice na\"{i}vely makes the problem worse, not better! Why? Because now the contact terms of the ladders, for example $\mathcal{L}_5$, which have only 3 degrees of freedom, have support on {\it five} cuts between those of (\ref{actual_w5_cuts_b}) and (\ref{actual_w5_cuts_c})---three and two, respectively. Nevertheless, it can be verified by direct computation that constraining the contact terms (\ref{wedge_int_example}) to vanish on the three cuts (\ref{actual_w5_cuts_b}), these integrals automatically vanish on the two additional cuts with the topology (\ref{actual_w5_cuts_c}). 

It is not hard to see that the problem we found here was unavoidable: there is no choice of cuts capable of defining the $\mathcal{W}_5$ integral which do not na\"{i}vely involve more constraints (on the contact terms of the ladders) than there exist available degrees of freedom. The cuts described above do enjoy the requisite magic, but we do not see why this {\it had} to work. 

As described in \mbox{section \ref{subsec:generalities_of_prescriptive_unitarity}}, if there had not been a solution to this problem, it would not have fundamentally spoiled our ability to write a closed formula for three loop amplitudes; it would have just prevented this from being a {\it prescriptive} representation. Suppose for example that we had chosen the `obvious' cuts to define $\mathcal{W}_5$ given in (\ref{w5_cuts_b}). The fact that ladder integrals including $\mathcal{L}_5$ could not be made to vanish on all four of the cuts in (\ref{w5_cuts_b}) would have meant that many coefficients from these higher terms would contribute to the one cut (of four) on which the integrals could not be made to vanish. Thus, the coefficient of this part of the $\mathcal{W}_5$ integrals in our basis could not be just `the corresponding cut in field theory', but the difference between the right answer in field theory and all the terms that pollute it. This would be a very-close-to prescriptive representation, but not strictly so. 

Clearly, this kind of tension should become more common at higher loop orders, and it would be very interesting to know if prescriptive representations continue to exist. We expect that this tension is avoidable through two loops in more general quantum field theories, but revisiting this story for more general theories at three loops would also be interesting.

\vspace{-0pt}\subsection{General Aspects of the Prescriptive Representation at Three Loops}\label{subsec:discussion_of_result}\vspace{-2pt}

The local integrand representation of three loop amplitude integrands in planar SYM derived here should be considered as an illustration of applying the prescriptive approach to generalized unitarity---well beyond the reach of earlier methods. Indeed, prior to this work, the only known expressions valid for all multiplicity were for MHV amplitudes, \cite{ArkaniHamed:2010gh}---a formula which was obtained essentially by guessing and comparing against the results of the loop-level BCFW recursion relations \cite{ArkaniHamed:2010kv}. The strategy we describe here seems much more general, explaining (to some extent) the surprising simplicity of loop amplitude integrands when expressed in terms of `pure' (or close-to-pure) local Feynman integrals as noticed in \mbox{ref.\ \cite{ArkaniHamed:2010gh}}. Ultimately, we seek a representation of loop amplitudes at the integrand level for which there is minimal cancellation between terms. Matching singularities of field theory one-by-one seems exactly in line with this goal.

While the specific result described here has virtually no relevance to the pressing computations needed for colliders, and only limited interest to even those researchers studying planar SYM, it represents a watershed of new theoretical data in which surprising features may be found. And because the formula (\ref{general_three_loop_rep_body}), when reinterpreted using on-shell functions of pure Yang-Mills represents a correct (albeit small) part of those amplitudes, the lessons we learn from this toy model have much broader implications for perturbation theory. For this reason, we would like to address some of the interesting aspects of these amplitudes, how this representation compares with others, and may be refined or recast to better expose different aspects of interest.

\vspace{-0pt}\subsubsection*{Potential for Specialization and Simplification}\label{subsubsec:possible_simplifications}\vspace{-2pt}

As described at the end of \mbox{section \ref{subsec:two_loop_unitarity}}, our construction of three loop integrands was (perhaps excessively) indifferent to the possible simplifications that arise for low multiplicity or for amplitudes with fixed N$^k$MHV-degree. Although our representation (\ref{general_three_loop_rep_body}) is arguably compact and general, it may not be the best representation for special classes of interesting amplitudes. 

One illustration of this would be a comparison with the only previously known all-multiplicity formula, for MHV amplitudes, as described in \mbox{ref.\ \cite{ArkaniHamed:2010gh}}. Using the notation here, that result was given as:
\vspace{-10pt}\eq{\mathcal{A}_n^{\text{MHV}}\bigger{=}\bigger{\displaystyle\sum_{\substack{\\[-5pt]\scalebox{0.45}{$a\!\leq\!b\!<\!c\!\leq\!$}\\
\scalebox{0.45}{$d\!<\!e\!\leq\!f\!<\!a$}
}}}\fwbox{100pt}{\fig{-47.3pt}{1}{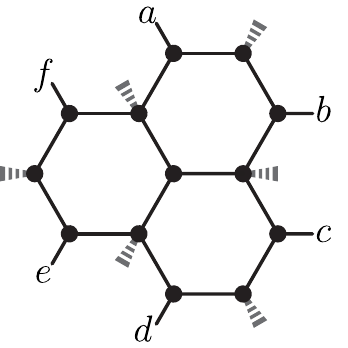}}\bigger{+}\bigger{\displaystyle\sum_{\substack{\\[-5pt]\scalebox{0.45}{$a\!\leq\!b\!<\!c\!<\!$}\\
\scalebox{0.45}{$d\!\leq\!e\!<\!f\!<\!a$}
}}}\fwbox{120pt}{\fig{-34.76pt}{1}{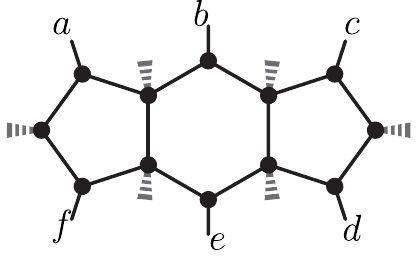}}\vspace{-5pt}\label{olde_mhv_three_loop}}
We refer the reader to \mbox{ref.\ \cite{ArkaniHamed:2010gh}} for a detailed description of these summands and the definitions of the tensor numerators defined for each integral. This representation is not incredibly different from the general expression valid for all N$^k$MHV amplitudes in (\ref{general_three_loop_rep_body}). Among the most obvious differences is the fact that there is no reference to on-shell functions as coefficients. This is explained by the fact that {\it all} (non-vanishing) leading singularities of planar MHV amplitudes are identical and equal to the tree amplitude---which has been factored out in (\ref{olde_mhv_three_loop}). 

Another salient distinction is that not all possible leg distributions are allowed for the integrands appearing in (\ref{olde_mhv_three_loop}). While many of the topologies from \mbox{Table \ref{table_of_integrals}} are included, only those involving many, specifically-placed massless legs are used. The reason for this is related to the fact that for any fixed N$^k$MHV degree not all on-shell functions appearing as coefficients may be non-vanishing. This is especially true for small $k$. To understand this, we should note that the N$^k$MHV-degree of an on-shell function corresponding to a graph $\Gamma$ involving amplitudes indexed by $v$ and $n_I$ internal lines is,
\eq{k_\Gamma=\sum_{v}k_v+2L-(4L-n_I)\,,\label{equation_for_k_of_graph}}
where $k_v$ is the N$^k$MHV-degree of each amplitude appearing in a corner. For a (non-composite) leading singularity, $(4L\mi n_I)\!=\!0$, and so in order for a on-shell function to be relevant to an MHV amplitude at three loops, $k_\Gamma\!=\!0\!=\sum_{v}k_v\!+\!6$. Because the only amplitudes for which $k\!<\!0$ are for three-point $\overline{\text{MHV}}$ amplitudes (for which $k\!=\!\mi1$), this is only possible if the on-shell function involves exactly $6$ such vertices, with all other amplitudes in the diagram being MHV, with $k\!=\!0$. This is the explanation for why each term in (\ref{olde_mhv_three_loop}) involves (generally) six massless legs, and why these integrals were drawn with empty three-point vertices at each vertex involving a massless leg in the work of \mbox{ref.\ \cite{ArkaniHamed:2010gh}}. Thus, the terms in (\ref{olde_mhv_three_loop}) almost exactly reflect the integrals with non-vanishing coefficients---all of which are equal to the MHV tree amplitude.\footnote{There is a curious exception for the wheel integral terms in (\ref{olde_mhv_three_loop}) within the topology $\mathcal{W}_8$: these integrals do not support MHV leading singularities in general. As such, we expect that there is unnecessary cancellation arising in the representation (\ref{olde_mhv_three_loop}), rendering it non-prescriptive.}

For N$^k$MHV amplitudes with $k\!<\!6$, such a specialization is always possible, as not all the cut topologies described in \mbox{Appendix \ref{appendix:details_three_loop_result}} have support in general. However, excluding some topologies comes at the cost of enumerating cases, which we expect will tend to introduce more complexity than would be gained. An exception may be the case of $k\!=\!1$, for which a restricted formulation of our general result may prove compact enough to be independently interesting. This is because NMHV amplitudes always support leading singularities,\footnote{This is manifestly true through three loops, but we expect it to be true more generally due to the existence of a `$d\!\log$' representation from BCFW recursion \cite{ArkaniHamed:2012nw}} and these residues are always simple `$R$-invariants'. Thus, we expect a representation exists for which no sub-leading cuts are required as coefficients. 

\vspace{-4pt}\subsubsection*{Composite Residues and (Exponentiation of) Infrared Divergences}\label{subsubsec:divergences_at_three_loops}\vspace{-2pt}

Another fruitful refinement of the general result may be to incorporate composite residues in order to expose the structure of infrared divergences. Our choice to {\it not} partition our representation (\ref{general_three_loop_rep_body}) according to finite and divergent parts does result in a more compact representation---as we saw also at two loops in \mbox{section \ref{subsec:two_loop_unitarity}}. Because the infrared divergences of loop amplitudes are always associated with soft-collinear regions in loop-momentum space, they should directly correspond to particular composite residues. And the structure of these divergences should roughly exponentiate as described by the BDS ansatz \cite{Bern:2005iz}. 

Besides avoiding an explosion of cases to consider and fixing integrals with soft-collinear (composite) residues separately from those without them, the principal reason why we did not do this here is that we do not understand how. We do not sufficiently understand the infrared divergences of the wheel-topology integrals to represent amplitudes in a way which suggests that these divergences exponentiate from lower loop-orders. We do not believe that there is any fundamental obstruction to doing so, but we leave the construction of such a representation to future research.

\vspace{-6pt}\subsubsection*{Transcendentality at Three Loops: Iterated and Elliptic Integrals}\label{subsubsec:elliptics_and_k3}\vspace{-2pt}
One final aspect of loop amplitudes at three loops worth mentioning involves the appearance of (potentially) non-polylogarithmic functions, including elliptic functions of various kinds. These contributions are intensely interesting, as our understanding of them is dramatically weaker than the purely polylogarithmic transcendental functions. As such, the necessity of this more general class of functions directly challenges our understanding, making new examples in which to study them valuable. 

The most concrete, unavoidable place where elliptic integrals arise in planar SYM is at two loops. As noted in \mbox{ref.\ \cite{CaronHuot:2012ab}}, the double-box integral involving all massive corners and at least one leg on each side of the middle propagator has no residues with maximal co-dimension---no leading singularities. This arrangement first arises for ten particles, 
\vspace{-10pt}\eq{\fig{-34.76pt}{1.36}{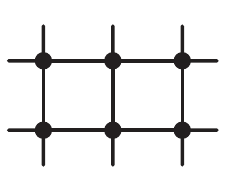}\vspace{-10pt}\label{ten_point_elliptic_double_box}}
and there is a strong argument why this integral is not an artifact of the representation: there exists an all-scalar component of the N$^3$MHV superamplitude for which the entire amplitude is given by just this integral (\ref{ten_point_elliptic_double_box}). (See \mbox{ref.\ \cite{Bourjaily:2015jna}} for details.)

It is easy to verify that the scalar loop integral (\ref{ten_point_elliptic_double_box}), on its co-dimension 7 residue cutting all propagators, results in a one-form on loop momentum space of the form of an elliptic integral. And it is not hard to be convinced that this is not an artifact: even when expressed as a four-fold integral over Feynman parameters, it has no co-dimension four residues, implying that it cannot be expressed as an iterated `$d\!\log$' integral by any change of variables. The most clear conclusion is that amplitudes even in planar SYM require a broader definition of transcendentality (not merely via polylogarithmic functions). 

Whether or not some scattering amplitudes {\it must} be polylogarithmic to all orders---even for restricted N$^k$MHV-degrees---has long been the subject of speculation (see {\it e.g.}\ \cite{Arkani-Hamed:2014via}). It would be beyond the scope of our present discussion to revisit these issues here, but we would like to point out that there is at least some evidence that even four-particle MHV amplitudes cannot be expressed {\it using local integrals} in polylogarithmic terms starting at eight loops \cite{Bourjaily:2015bpz}; and this fact has images at lower loops and higher multiplicity---including the ten particle example mentioned above at two loops. 

While we do not want to speculate much on the implications for transcendentality of our three loop representation, there are some intriguing aspects that deserve further investigation. The most obvious new class of non-polylogarithmic functions that arise at three loops (not merely by having a sub integral corresponding to (\ref{ten_point_elliptic_double_box})) is for the generally massive instance of the ladder $\mathcal{L}_1$:
\vspace{-12pt}\eq{\fig{-34.76pt}{1.36}{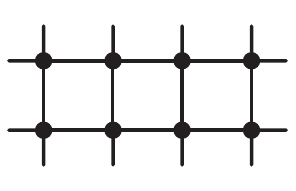}\vspace{-14pt}\label{twelve_point_triple_box}}
Cutting all ten propagators of this integral results in a two-form on loop momenta parameterized $(x,y)$ of the form,
\vspace{-12pt}\eq{\fig{-34.76pt}{1.36}{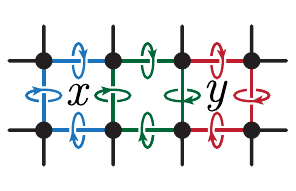}=\int\!\!\frac{dx\,dy}{\sqrt{Q(x,y)}}\,,\vspace{-10pt}\label{double_elliptic_integral}}
where $Q(x,y)$ is an irreducible quartic polynomial in each variable. The precise implications of this observation for the transcendental structure of the loop integral (\ref{twelve_point_triple_box}) remains unclear, but intensely interesting. The coefficient of $\mathcal{L}_1$ has support on this co-dimension ten residue for twelve particles only for N$^4$MHV amplitudes. Indeed, there exists a scalar component for which the amplitude precisely takes the form of (\ref{double_elliptic_integral}) on this cut---strongly suggesting that these kinds of integrals are unavoidable parts of loop amplitudes. 

The final example of transcendental novelty at three loops arises in the case of the wheel integral $\mathcal{W}_1$, again for a generally massive distribution of external legs:
\vspace{-9pt}\eq{\fig{-34.76pt}{1.36}{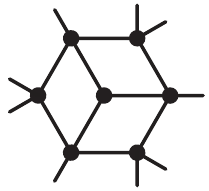}\vspace{-9pt}\label{nine_point_wheel}}
This new class of integral first arises for nine particles with coefficients supported for N$^2$MHV or N$^3$MHV amplitudes (which are parity conjugate). Unlike the case above, cutting all nine propagators in (\ref{nine_point_wheel}) results in a strictly rational form on the three remaining degrees of freedom. This rational three-form, however, does support co-dimension one residues of the same form as in (\ref{double_elliptic_integral}). (This does not occur for fewer than nine external legs distributed as in (\ref{nine_point_wheel}).) What this implies about the integrated form of the Feynman integral (\ref{nine_point_wheel}), and the implications of this integral for scattering amplitudes, however, remains unclear. In particular, while the three-form resulting from cutting all nine propagators in (\ref{nine_point_wheel}) has a co-dimension one residue of an elliptic type, there are no co-dimension ten cuts of a nine point amplitude which have this form. (There are no $\eta$-components of the superfunctions on which this cut would be non-vanishing.) Thus, even if the integral $\mathcal{W}_1$ were not polylogarithmic, this would not imply that nine-point {\it amplitudes} are so: it may merely represent an artifact of the local integrand representation. 
 
The situation described above is reminiscent of the case of N$^2$MHV amplitudes at two loops. While it is easy to prove on general grounds that no N$^2$MHV amplitude has support on an elliptic cut at two loops, this fact need not be made manifest term-by-term in an integrand representation. Indeed, while a (prescriptive) representation which makes manifest the non-ellipticity of these amplitudes at two loops does exist, neither the representation in this work nor that in \mbox{ref.\ \cite{Bourjaily:2015jna}} makes this fact manifest: the basis of integrals used has many term-wise elliptic contributions. 

Whether such term-wise ellipticity is an artifact of the representation, or a necessary consequence of using a local loop expansion remains to be seen. Further investigations of these properties of loop amplitudes would be worthwhile.

\newpage
\vspace{-12pt}\section{Prescriptive Unitarity at One Loop For General Theories}\label{sec:one_loop_revisited}\vspace{-8pt}
In our review of one loop unitarity in \mbox{section \ref{subsec:one_loop_unitarity}}, we concluded with a perhaps perplexing comment about the difficulty of employing this approach to SYM. In this section, we clarify this comment, and show how a prescriptive approach can be employed---at the cost of making the power counting of the theory non-manifest. In the process, we will illustrate how the approach we describe here compares with the more traditional approach for theories with more general power counting. For the sake of illustration, we will continue to discuss theories in (strictly) four dimensions; as such, our examples here will be limited to the cut-constructible parts of dimensionally-regulated theories in four dimensions.

Recall from our discussion in \mbox{section \ref{subsec:one_loop_unitarity}} that an over-complete basis of integrals for a theory with ultraviolet behavior dictated by $\sim\!1/(\ell^2)^4$ would be:
\vspace{-10pt}\eq{\left\{\raisebox{34pt}{$\!\!$}\right.\fwbox{55pt}{\fig{-34.76pt}{1.36}{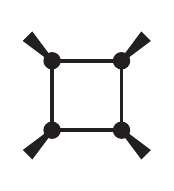}},\fwbox{75pt}{\fig{-34.76pt}{1.36}{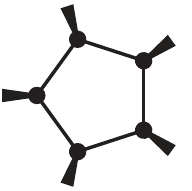}}\!\!\left.\raisebox{34pt}{$\!\!$}\right\}\label{minus_four_power_counting_basis}\vspace{-10pt}}
This basis, while complete, is over-complete. A choice of independent parity-odd pentagons must be made in order to even define coefficients $c_k$ of an amplitude's integrand.

Conveniently, as we saw in \mbox{section \ref{subsec:one_loop_unitarity}}, we may always without loss of generality expand the identity polynomial in terms of inverse propagators, which means that the power counting of a theory need only represent a lower bound: (at the cost of introducing an arbitrary scale into the representation), we may always consider loop integrands in SYM to be expanded into integrands with the power counting of $\sim\!1/(\ell^2)^3$. Considering loop integrands in SYM to have this power counting is at worst a bad idea (we will see that it is not); for a more general quantum field theory, this is a necessary case for us to consider. 

Therefore, let us now construct a basis of one loop integrands which scale asymptotically as $\sim\!1/(\ell^2)^3$ for large loop momenta. Clearly, any integrand in our basis must have at least three propagators; and---as always---Passarino-Veltman reduction allows us to focus our attention to those with at most five external propagators. Thus, we may na\"{i}evely have an over-complete basis of scalar triangles, tensor boxes, and pentagons involving two powers of inverse propagators in their numerator.

Box integrals with a single inverse propagator in the numerator have 6 degrees of freedom, which cleanly separate (using the ``chiral'' basis of (\ref{chiral_pentagon_basis})) into 2 non-contact (`chiral') degrees of freedom, and 4 contact terms. And following the argument at the end of \mbox{section \ref{subsec:two_loop_unitarity}}, it is easy to see that pentagon integrals with two inverse propagators can be entirely decomposed into contact terms: $20\!=\!\binom{5}{1}\!\times\!2\pl\binom{5}{2}\!\times\!1$. Thus, a complete basis of integrals consistent with this power counting would consist of just `chiral' boxes, with two (non-contact) degrees of freedom each, and scalar triangles with one degree of freedom each:
\vspace{-10pt}\eq{\left\{\raisebox{34pt}{$\!\!$}\right.\fwbox{65pt}{\fig{-34.76pt}{1.36}{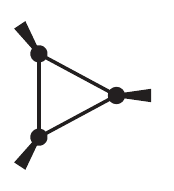}},\fwbox{60pt}{\fig{-34.76pt}{1.36}{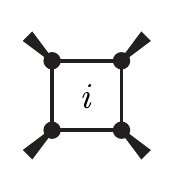}}\left.\raisebox{34pt}{$\!\!$}\right\}\label{minus_three_power_counting_basis}\vspace{-10pt}}
Conveniently, it turns out that (for $n\!>\!4$), this basis is not over-complete. To see this, we imagine combining all integrals over a common denominator and observe that this power counting requires $r\!=\!n\mi3$ powers of inverse propagators in the numerator; according to (\ref{dimension_counting_by_power}), this space of numerators has dimension $\binom{n}{4}\pl\binom{n+1}{4}$; and the basis (\ref{minus_three_power_counting_basis}) consists of $\binom{n}{4}\!\times\!2\pl\binom{n}{3}\!\times\!1$, which matches the correct counting (for $n\!>\!4$).  

In terms of this basis, a prescriptive representation is easy to construct. The triangle integrals have only one degree of freedom, and therefore should be defined in order to match field theory at an arbitrary point along the triple-cut involving the three propagators. The 2 non-contact degrees of freedom of each chiral box can be chosen to match field theory on the two box-type leading singularities (see \ref{triangle_to_box})), and their 4 contact-term degrees of freedom should be chosen to vanish at the arbitrary points where the triangles are defined. Thus, the cuts which define the integrals in (\ref{minus_three_power_counting_basis}) together with their coefficients would be, respectively:
\vspace{-5pt}\eq{\left\{\raisebox{40pt}{$\!\!$}\right.\fwbox{65pt}{\fig{-34.76pt}{1.36}{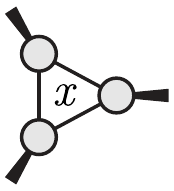}}\,\,,\!\!\fwbox{60pt}{\fig{-34.76pt}{1.36}{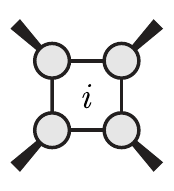}}\left.\raisebox{40pt}{$\!\!$}\right\}\label{minus_three_power_counting_coefficients}\vspace{-5pt}}

This prescriptive representation of one loop amplitudes with worse-than-SYM power-counting is in fact very close to the prescriptive representation derived in \mbox{ref.\ \cite{Bourjaily:2013mma}}. The principle distinction between the discussion above and the result described in \mbox{ref.\ \cite{Bourjaily:2013mma}} is that composite residues were used to fix the coefficients of the triangle integrals (instead of arbitrary points here). Also, in \mbox{ref.\ \cite{Bourjaily:2013mma}}, spurious propagators were included in every integral of the basis---trading the wrong power counting for non-manifest dual conformal invariance of the result. This may or may not be the best representation of integrands in SYM---as the power-counting of the theory is rendered non-manifest; but it does correctly capture any quantum field theory bounded by this degree of divergence in the ultraviolet.  

The worst power counting of any four-dimensional quantum field theory without tadpoles would be $\sim\!1/(\ell^2)^2$. As before, we may without any loss of generality consider any theory with better power counting to be included in this case. As such, it captures the cut-constructible part of any quantum field theory at one loop (including those with better ultraviolet behavior). 

Following the logic which should now be familiar, it is clear that we may expand any integral into those with two, three, four, or five propagators, with $1$, $6$, $20$, and $50$ degrees of freedom in the numerator for each. The $6$ degrees of freedom of the general triangle integral split into $3$ non-contact terms and $3$ contact terms according to the basis (\ref{3_2_pentagon_basis}), and the 20 degrees of freedom of the boxes split into $2$ non-contact and $18\!=\!\binom{4}{1}\!\times\!3\pl\binom{4}{2}\!\times\!1$ contact degrees of freedom. Following the same discussion as for three loops, the $50$ degrees of freedom of a degree-three tensor product of inverse propagators for an integral with $5$ external propagators can be entirely decomposed into contact terms: $50\!=\!\binom{5}{1}\!\times\!2\pl\binom{5}{2}\!\times\!3\pl\binom{5}{3}\!\times\!1$. Thus a complete basis of loop integrands with this ultraviolet behavior can be represented in terms of:
\vspace{-10pt}\eq{\left\{\raisebox{34pt}{$\!\!$}\right.\fwbox{74pt}{\fig{-34.76pt}{1.36}{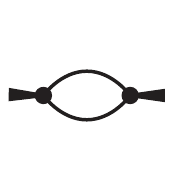}},\fwbox{65pt}{\fig{-34.76pt}{1.36}{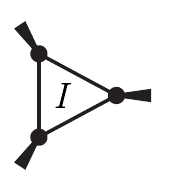}},\fwbox{60pt}{\fig{-34.76pt}{1.36}{chiral_box}}\left.\raisebox{34pt}{$\!\!$}\right\}\label{minus_two_power_counting_basis}\vspace{-10pt}}
Here, the triangle integrals have 3 non-contact degrees of freedom in their numerator, indexed by $I\!\in\!\{1,2,3\}$. Conveniently, as with the case of $1/(\ell^2)^3$ power-counting (and in contrast to the case of $1/(\ell^2)^4$), the basis of integrals in (\ref{minus_two_power_counting_basis}) is not over-complete:
\eq{\binom{n+1}{4}+\binom{n+2}{4}=\binom{n}{2}\!\times\!1+\binom{n}{3}\!\times\!3+\binom{n}{4}\!\times\!2\,.}
(Unlike before, the independence of this basis holds for any $n$---including $n\!=\!4$.)

From this basis, a prescriptive representation is easy to construct. The coefficients of the bubbles are fixed to match field theory at some two-parameter point $\vec{x}$; and the $\binom{3}{1}$ contact terms of the triangles and $\binom{4}{2}$ double-contact terms of the chiral boxes are determined by the requirement that these integrals vanish at this point. There is something new for the cuts used to define the non-contact term degrees of freedom of the triangle integrals: all the cuts which define these integrals (and also fix their coefficients in the representation) have the same topology. This is not a problem: we simply choose any three points $x_I$, with $I\!\in\!\{1,2,3\}$ along the cut. The non-contact-term degrees of freedom of the chiral boxes are, as before, determined by the leading singularities of (\ref{triangle_to_box}). 

Thus, the basis of integrals in (\ref{minus_two_power_counting_basis}) can be defined in terms of the following cuts, which also determine their coefficients in the representation of the amplitude:
\vspace{-7pt}\eq{\left\{\raisebox{34pt}{$\!\!$}\right.\fwbox{74pt}{\!\!\fig{-34.76pt}{1.36}{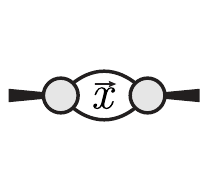}},\fwbox{65pt}{\!\fig{-34.76pt}{1.36}{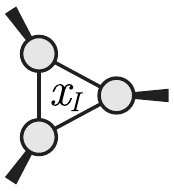}},\fwbox{60pt}{\fig{-34.76pt}{1.36}{chiral_box_cut}}\left.\raisebox{34pt}{$\!\!$}\right\}\label{minus_two_power_counting_coefficients}\vspace{-7pt}}

It is worthwhile to compare this approach with what is ordinarily done using OPP \cite{Ossola:2006us}. Ordinarily, the coefficients of, {\it e.g.}, the triangle integrals are determined by sufficient evaluation at a sufficient number of points along their maximal cut (the residue cutting all three propagators); because box integrals (whether scalar or tensor) do not vanish on these triple-cuts, the coefficients of the triangles are the difference between the `right answer in field theory' (the on-shell function, evaluated at these points) and the sum of box coefficients which `pollute' this cut. What we are doing here amounts to a reorganization of the terms that result (and a better strategy to find them). Even for the tensor triangles integrals required for a theory with $\sim\!1/(\ell^2)^2$ power counting, we do require evaluations of triangle integrals at several points (namely, three), matched by field theory at these points. Rather than using `pure' box integrals, however, we are {\it defining} the `chiral' boxes in our basis (\ref{minus_two_power_counting_basis}) to vanish at these points along the triple cuts. This operation requires that the `box' integrals in our basis include contact-terms with the topology of triangles. Thus, the triangle integrals in our basis (\ref{minus_two_power_counting_basis}) include also contact-term degrees of freedom with the topology of bubbles, and the boxes include both triangle- and bubble-topology contact terms. 

This reorganization is not extremely different from what is ordinarily achieved using generalized unitarity at one loop. However, we hope that our illustrations at two and three loops (even for the simple case of planar SYM) demonstrate the advantages of organizing the basis according to the prescriptive approach outlined in \mbox{section \ref{subsec:generalities_of_prescriptive_unitarity}}.

Let us conclude with a general discussion of the division of integrand numerators into non-contact terms and contact terms relevant to theories with more general power counting and also to SYM at higher loop orders. We have seen many explicit examples of this already, but it is worthwhile to notice some of the general trends. The separation of the degrees of freedom for the numerator of a loop integrand involving $r$ powers of inverse propagators which includes $n_{\text{ext}}$ external propagators for that loop momentum is given in \mbox{Table \ref{power_counting_table}}. Finally, it is worth mentioning that there may be ambiguity in the identification of `external' propagators for a non-planar graph; whenever this occurs, the non-uniqueness implies identities among different bases for the numerators, reducing the overall counting relative to \mbox{Table \ref{power_counting_table}}.

\begin{table}[b]\eq{\begin{array}{|@{$\,$}c@{$\,$}|@{$$}c@{$\,$}|@{$\,$}c@{$\,$}|@{$\,$}c@{$\,$}|@{$\,$}c@{$\,$}|@{$\,$}c@{$\,$}|@{$\,$}c@{$\,$}|}\multicolumn{7}{c}{}\\[-40pt]\multicolumn{1}{@{}c@{$\,$}}{}&\multicolumn{6}{@{$$}c@{$\,$}}{r}\\\cline{2-7}\multicolumn{1}{@{}c@{$\,$}|@{$$}}{n_{\text{ext}}}&\hspace{2pt}0&1&2&3&4&5\\
\hline0&\hspace{2pt}{\color{hblue}1}&{\color{hblue}6}{\color{dred}\pl0}&{\color{hblue}20}{\color{dred}\pl0\phantom{1}}&{\color{hblue}50}{\color{dred}\pl0\phantom{1}}&{\color{hblue}105}{\color{dred}\pl0\phantom{15}}&{\color{hblue}196}{\color{dred}\pl0\phantom{16}}\\\hline
1&\hspace{2pt}{\color{hblue}1}&{\color{hblue}5}{\color{dred}\pl1}&{\color{hblue}14}{\color{dred}\pl6\phantom{1}}&{\color{hblue}30}{\color{dred}\pl20}&{\color{hblue}\phantom{1}55}{\color{dred}\pl50\phantom{1}}&{\color{hblue}\phantom{1}91}{\color{dred}\pl105\phantom{}}\\\hline
2&\hspace{2pt}{\color{hblue}1}&{\color{hblue}4}{\color{dred}\pl2}&{\color{hblue}\phantom{1}9}{\color{dred}\pl11}&{\color{hblue}16}{\color{dred}\pl34}&{\color{hblue}\phantom{1}25}{\color{dred}\pl80\phantom{1}}&{\color{hblue}\phantom{1}36}{\color{dred}\pl160\phantom{}}\\\hline
3&\hspace{2pt}{\color{hblue}1}&{\color{hblue}3}{\color{dred}\pl3}&{\color{hblue}\phantom{1}5}{\color{dred}\pl15}&{\color{hblue}\phantom{1}7}{\color{dred}\pl43}&{\color{hblue}\phantom{01}9}{\color{dred}\pl96\phantom{1}}&{\color{hblue}\phantom{1}11}{\color{dred}\pl185\phantom{}}\\\hline
4&\hspace{2pt}{\color{hblue}1}&{\color{hblue}2}{\color{dred}\pl4}&{\color{hblue}\phantom{1}2}{\color{dred}\pl18}&{\color{hblue}\phantom{1}2}{\color{dred}\pl48}&{\color{hblue}\phantom{10}2}{\color{dred}\pl103}&{\color{hblue}\phantom{01}2}{\color{dred}\pl194\phantom{}}\\\hline
5&\hspace{2pt}{\color{hblue}1}&{\color{hblue}1}{\color{dred}\pl5}&{\color{hblue}\phantom{1}0}{\color{dred}\pl20}&{\color{hblue}\phantom{1}0}{\color{dred}\pl50}&{\color{hblue}\phantom{15}0}{\color{dred}\pl105}&{\color{hblue}\phantom{19}0}{\color{dred}\pl196}\\\hline\multicolumn{7}{c}{}\\[-26pt]
\end{array}\nonumber}
\caption{Division of numerator degrees of freedom consisting of $r$ powers of inverse propagators into {\color{hblue}non-contact} vs.\ {\color{hred}contact} terms for integrands involving $n_{\text{ext}}$ external propagators.\label{power_counting_table}}\vspace{-42pt}\end{table}

\newpage
\vspace{-12pt}\section{Conclusions and Future Directions}\label{sec:conclusions}\vspace{-8pt}
In this work we have described a new strategy for implementing generalized unitarity to (re)construct local integrand representations of scattering amplitudes in any quantum field theory. We call this approach {\it prescriptive} because the absence of linear algebra in our approach results in closed-form representations of amplitudes. We described this approach in considerable generality, using applications to planar SYM for the sake of illustration. These applications include the first determination of all multiplicity, all N$^k$MHV scattering amplitudes in this theory through three loops---a considerable advance in theoretical data. 

Despite our use of planar SYM for illustration, we are optimistic that this strategy will prove useful more generally. In particular, it would be very worthwhile to seek prescriptive representations of loop amplitudes for non-planar theories, theories without supersymmetry, and for theories defined in arbitrary (or dimensionally-regulated) dimensions. Nothing about our strategy requires any of these simplifications, but it remains to be demonstrated that prescriptive representations exist more generally. This is especially the case because, as we have seen, the mere existence of a prescriptive representation at three loops for planar SYM required non-trivial magic to work. This tension demonstrates the non-triviality of the existence of strictly prescriptive representations of loop amplitudes, and its resolution represents evidence that something special is at work which remains to be fully understood. 

We have uncovered new non-polylogarithmic structures of amplitudes at three loops, the implications of which can be better understood now that entire loop amplitude integrands are known. The consequences of these structures for the symbolic bootstrap program (see {\it e.g.}\ \cite{Dixon:2011pw,Dixon:2011nj,Dixon:2013eka,Dixon:2014iba,Dixon:2015iva,Caron-Huot:2016owq}) would be fruitful to explore. 

There are many interesting roads ahead for further research. Beyond the application of these ideas to more general theories, we expect there to be illuminating refinements and reformulations of our results for planar SYM. In particular, it would be worthwhile to find simpler representations for low multiplicity or low N$^k$MHV-degree, and interesting to make the (exponentiation of) infrared divergences of scattering amplitudes manifest at the integrand level.

\vspace{-12pt}\section*{Acknowledgements}\vspace{-10pt}
We gratefully acknowledge Zvi Bern for insightful comments on the manuscript and thank JJ Carrasco and JJ Stankowicz for helpful discussions. This work was supported in part by the National Science Foundation under Grant No.\ NSF PHY-1125915; by the Danish National Research Foundation (DNRF91) and a grant from the Villum Fonden (JLB); and by a grant from the Gordon and Betty Moore Foundation and by DOE Grant No.\ DE-SC0011632 (EH).

\newpage\appendix
\vspace{-6pt}\section{Explicit Contributions to Three Loop Amplitudes}\label{appendix:details_three_loop_result}\vspace{-6pt}

In this Appendix, we enumerate the cut conditions which define the non-contact-term degrees of freedom of every integral appearing in the three loop basis described in \mbox{Table \ref{table_of_integrals}}. The contact-term degrees of freedom are always entirely fixed by the criteria that these integral vanish on all the cuts used to define other integrals in the basis. Because the resulting basis is diagonal in cuts, the coefficient of every integral is then fixed to be the corresponding field theory cut. Thus, we use these field theory cut pictures to represent how each of the non-contact degrees of freedom are fixed. 

\vspace{-0pt}\subsection{Detailed Description of Wheel Integrals: Defining Cuts/Coefficients}\label{subsec:wheel_integral_sequence}\vspace{-2pt}

\eq{\fwbox{0pt}{\fwboxL{430pt}{\begin{array}{@{}c@{}c@{}}
\fwbox{100pt}{\fig{-47.3pt}{1}{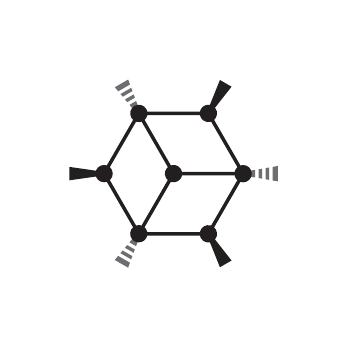}}\fwboxR{0pt}{\left\{\raisebox{34pt}{}\right.}&\hspace{-5.5pt}\fwbox{90pt}{\fig{-47.3pt}{1}{wheel_cut_1a}}\fwboxR{0pt}{\left.\raisebox{34pt}{}\right\}}\\[-19.5pt]
\fwboxL{100pt}{\text{{\footnotesize degrees of freedom:}}}&1
\end{array}}\fwboxR{0pt}{\raisebox{-3.pt}{$(\mathcal{W}_1)$}}}\nonumber}
\vspace{-20pt}
\eq{\fwbox{0pt}{\fwboxL{430pt}{\begin{array}{@{}c@{}@{}c@{}c@{}}\fwbox{100pt}{\fig{-47.3pt}{1}{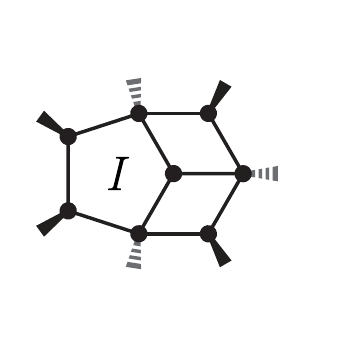}}\fwboxR{0pt}{\left\{\raisebox{40pt}{}\right.}&\hspace{-30pt}\fwbox{100pt}{\hspace{45pt}\fig{-47.3pt}{1}{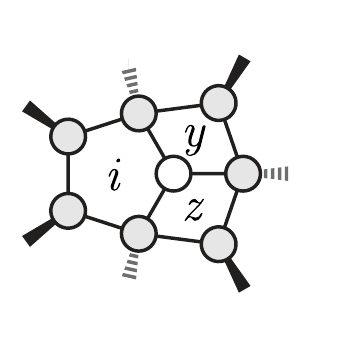}}&
\hspace{0pt}\fwbox{120pt}{\fig{-47.3pt}{1}{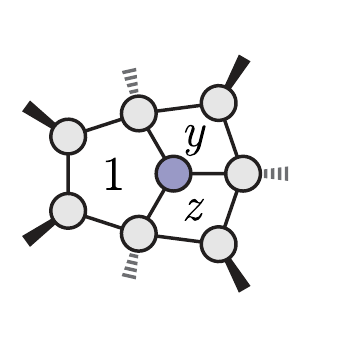}}\hspace{-10pt}\fwboxR{0pt}{\left.\raisebox{40pt}{}\right\}}\\[-14.5pt]
\fwboxL{100pt}{\text{{\footnotesize degrees of freedom:}}}&2&1
\end{array}}\fwboxR{0pt}{\raisebox{-0pt}{$(\mathcal{W}_2)$}}}\nonumber}
\vspace{-10pt}
\eq{\fwbox{0pt}{\fwboxL{430pt}{\begin{array}{@{}c@{}@{}c@{}c@{}}
\fwbox{100pt}{\fig{-47.3pt}{1}{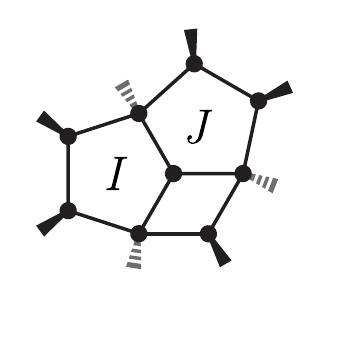}}\fwboxR{0pt}{\left\{\raisebox{40pt}{}\right.}&\hspace{-30pt}\fwbox{100pt}{\hspace{45pt}\fig{-47.3pt}{1}{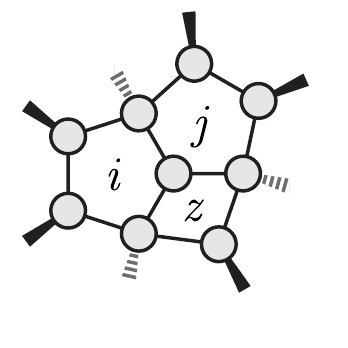}}&
\hspace{0pt}\fwbox{120pt}{\fig{-47.3pt}{1}{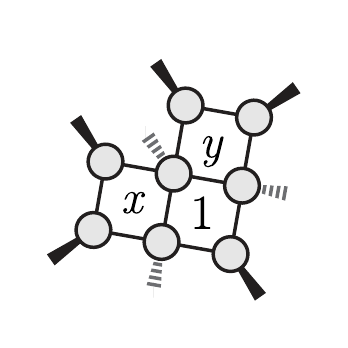}}\hspace{-10pt}\fwboxR{0pt}{\left.\raisebox{40pt}{}\right\}}\\[-14.5pt]
\fwboxL{100pt}{\text{{\footnotesize degrees of freedom:}}}&8&1
\end{array}}\fwboxR{0pt}{\raisebox{0pt}{$(\mathcal{W}_3)$}}}\nonumber}
\vspace{-15pt}
\eq{\fwbox{0pt}{\fwboxL{430pt}{\begin{array}{@{}c@{}c@{}}
\fwbox{100pt}{\fig{-47.3pt}{1}{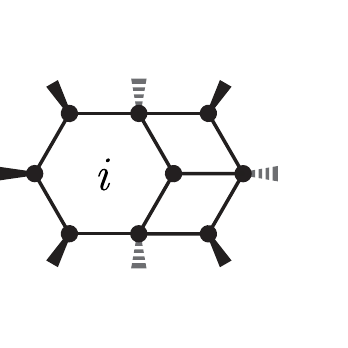}}\fwboxR{0pt}{\left\{\raisebox{34pt}{}\right.}&\hspace{5.5pt}\fwbox{80pt}{\fig{-47.3pt}{1}{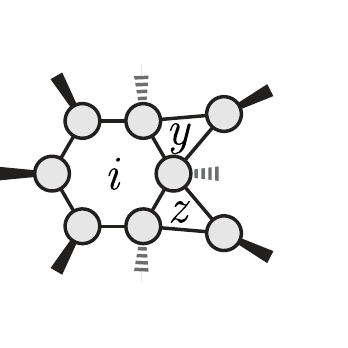}}\fwboxR{0pt}{\left.\raisebox{34pt}{}\right\}}\\[-15.5pt]
\fwboxL{100pt}{\text{{\footnotesize degrees of freedom:}}}&2
\end{array}}\fwboxR{0pt}{\raisebox{-0.5pt}{$(\mathcal{W}_4)$}}}\nonumber}
\vspace{-10pt}
\eq{\fwbox{0pt}{\fwboxL{430pt}{\begin{array}{@{}c@{}@{}c@{}c@{}c@{}}\fwbox{100pt}{\fig{-47.3pt}{1}{wheel_int_5}}\fwboxR{0pt}{\left\{\raisebox{40pt}{}\right.}&\hspace{-2.5pt}\fwbox{90pt}{\fig{-47.3pt}{1}{wheel_cut_5a}}&
\fwbox{90pt}{\fig{-47.3pt}{1}{wheel_cut_5b}}&\hspace{-15pt}\fwbox{90pt}{\fig{-47.3pt}{1}{wheel_cut_5c}}\fwboxR{0pt}{\left.\raisebox{40pt}{}\right\}}\\[-2.5pt]
\fwboxL{100pt}{\text{{\footnotesize degrees of freedom:}}}&15&3\!\times\!3&1\!\times\!3
\end{array}}\fwboxR{0pt}{\raisebox{6.25pt}{$(\mathcal{W}_5)$}}}\nonumber}
\vspace{-10pt}
\eq{\fwbox{0pt}{\fwboxL{430pt}{\begin{array}{@{}c@{}@{}c@{}c@{}}\fwbox{100pt}{\fig{-47.3pt}{1}{wheel_int_6}}\fwboxR{0pt}{\left\{\raisebox{40pt}{}\right.}&\hspace{0pt}\fwbox{100pt}{\fig{-47.3pt}{1}{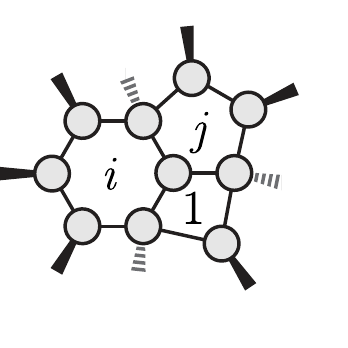}}&
\fwbox{100pt}{\fig{-47.3pt}{1}{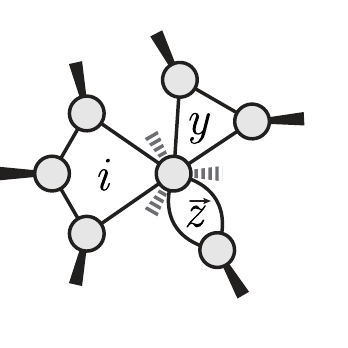}}\fwboxR{0pt}{\left.\raisebox{40pt}{}\right\}}\\[-15.5pt]
\fwboxL{100pt}{\text{{\footnotesize degrees of freedom:}}}&4&2
\end{array}}\fwboxR{0pt}{(\mathcal{W}_6)}}\nonumber}
\vspace{-10pt}
\eq{\fwbox{430pt}{\fwboxL{430pt}{\hspace{-2.5pt}\begin{array}{@{}c@{}@{}c@{}c@{}c@{}c@{}}\fwbox{100pt}{\fig{-47.3pt}{1}{wheel_int_7}}\hspace{-5pt}\fwboxR{0pt}{\left\{\raisebox{40pt}{}\right.}&\hspace{0.5pt}\fwbox{85pt}{\fig{-47.3pt}{1}{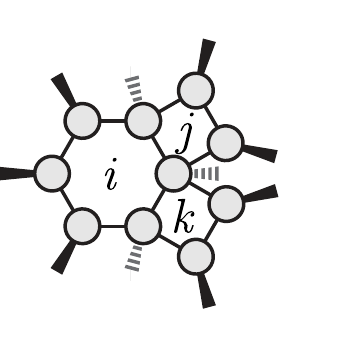}}&
\fwbox{80pt}{\fig{-47.3pt}{1}{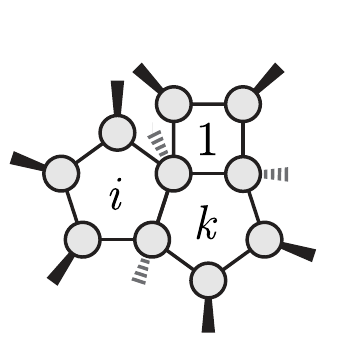}}&\hspace{-0pt}\fwbox{90pt}{\fig{-47.3pt}{1}{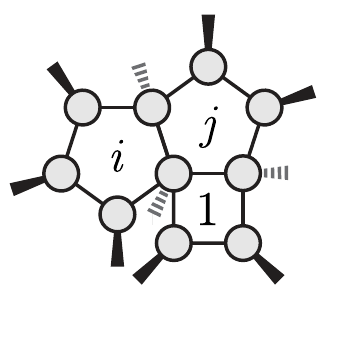}}&\hspace{-0pt}\fwbox{90pt}{\fig{-47.3pt}{1}{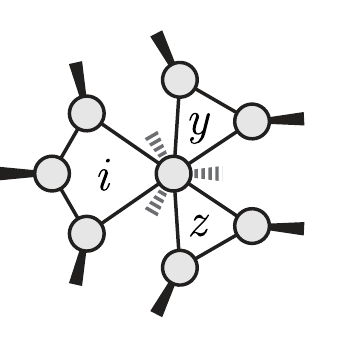}}\fwboxR{0pt}{\left.\raisebox{40pt}{}\right\}}\\[-10.5pt]
\fwboxL{100pt}{\text{{\footnotesize degrees of freedom:}}}&8&\hspace{-10pt}4&4&\hspace{-20pt}2
\end{array}}\fwboxR{0pt}{\raisebox{-44pt}{$(\mathcal{W}_7)$}}}\nonumber}
\vspace{-10pt}
\eq{\fwbox{0pt}{\fwboxL{430pt}{\begin{array}{@{}c@{}c@{}}\fwbox{100pt}{\fig{-47.3pt}{1}{wheel_int_8}}\hspace{10pt}\fwboxR{0pt}{\left\{\raisebox{42pt}{}\right.}\hspace{-5pt}&\hspace{-5.5pt}\fwbox{120pt}{\fig{-47.3pt}{1}{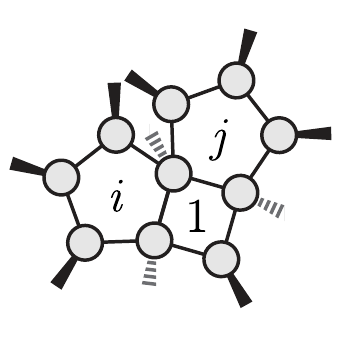}}\fwboxR{0pt}{\left.\raisebox{42pt}{}\right\}}\\[-15.5pt]
\fwboxL{100pt}{\text{{\footnotesize degrees of freedom:}}}&4
\end{array}}\fwboxR{0pt}{\raisebox{-0.5pt}{$(\mathcal{W}_8)$}}}\nonumber}
\vspace{-10pt}
\eq{\fwbox{0pt}{\fwboxL{430pt}{\begin{array}{@{}c@{}@{}c@{}c@{}}\fwbox{100pt}{\fig{-47.3pt}{1}{wheel_int_9}}\hspace{10pt}\fwboxR{0pt}{\left\{\raisebox{40pt}{}\right.}\hspace{-5pt}&\hspace{-0pt}\fwbox{100pt}{\fig{-47.3pt}{1}{wheel_cut_9a}}&
\fwbox{100pt}{\fig{-47.3pt}{1}{wheel_cut_9b}}\hspace{5pt}\fwboxR{0pt}{\left.\raisebox{40pt}{}\right\}}\\[-5.5pt]
\fwboxL{100pt}{\text{{\footnotesize degrees of freedom:}}}&8&4
\end{array}}\fwboxR{0pt}{\raisebox{4.5pt}{$(\mathcal{W}_9)$}}}\nonumber}
\vspace{-10pt}
\eq{\fwbox{0pt}{\fwboxL{430pt}{\begin{array}{@{}c@{}c@{}}\fwbox{100pt}{\fig{-47.3pt}{1}{wheel_int_10}}\hspace{10pt}\fwboxR{0pt}{\left\{\raisebox{42pt}{}\right.}\hspace{-5pt}&\hspace{2.5pt}\fwbox{105pt}{\fig{-47.3pt}{1}{wheel_cut_10a}}\fwboxR{0pt}{\left.\raisebox{42pt}{}\right\}}\\[-5.5pt]
\fwboxL{100pt}{\text{{\footnotesize degrees of freedom:}}}&8
\end{array}}\fwboxR{0pt}{\raisebox{4.15pt}{$(\mathcal{W}_{10})$}}}\nonumber}

Some clarifying comments are in order. For the amplitudes involved in on-shell function coefficients, those coloured in grey denote any N$^k$MHV-degree; this applies also to three-point amplitudes, except where specifically indicated. There are only two-three point amplitudes, coloured blue and white by convention to indicate MHV and $\overline{\text{MHV}}$, respectively. For $\mathcal{W}_2$, for example, we want to make clear that the counting of coefficients depends on the degree of the middle three-point amplitude. (The choice of how these are chosen is arbitrary, but must be made.) For $\mathcal{W}_3$, the fact that the middle three-point amplitude can have two possible MHV-degrees accounts for there being eight cuts with the first topology---indexed by $i,j$ and the middle three-point amplitude. This convention may seem to be in conflict with the counting for $\mathcal{W}_6$, but this is in fact accounted for by the choice of cut `$1$' for the box---which fixes the MHV-degree of the middle amplitude implicitly. 

Finally, as described in the body of the text (see \mbox{section \ref{subsubsec:very_carefully_chosen_cuts}}), the counting for $\mathcal{W}_5$ requires some explanation. There are na\"{i}vely 16 cuts with the first topology---indexed by $(i,j,k)$ and the degree of the middle amplitude; any 15 of these are independent and some (arbitrary) choice of which 15 must be made. For the second class of cuts, there are 4 possible cuts as drawn (all of which are independent, but we have chosen not to make use of this fact); as such, the `$3\times$' in the counting reflects this choice of 3 of the 4 possible cuts; for both the second and third topologies, `$\times3$' indicates cuts related by cyclic rotation. 

\vspace{-0pt}\subsection{Detailed Description of Ladder Integrals: Defining Cuts/Coefficients}\label{subsec:ladder_integral_sequence}\vspace{-2pt}

\vspace{-10pt}
\eq{\fwbox{0pt}{\fwboxL{430pt}{\begin{array}{@{}c@{}c@{}}\fwbox{100pt}{\fig{-34.76pt}{1}{ladder_int_1}}\hspace{10pt}\fwboxR{0pt}{\left\{\raisebox{29pt}{}\right.}\hspace{-15pt}&\hspace{0.5pt}\fwbox{115pt}{\fig{-34.76pt}{1}{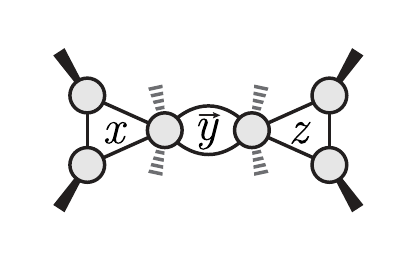}}\fwboxR{0pt}{\left.\raisebox{29pt}{}\right\}}\\[-15.5pt]
\fwboxL{100pt}{\text{{\footnotesize degrees of freedom:}}}&1
\end{array}}\fwboxR{0pt}{\raisebox{0.35pt}{$(\mathcal{L}_{1})$}}}\nonumber}
\vspace{-5pt}
\eq{\fwbox{0pt}{\fwboxL{430pt}{\begin{array}{@{}c@{}c@{}c@{}c@{}}\fwbox{100pt}{\fig{-34.76pt}{1}{ladder_int_2}}\hspace{10pt}\fwboxR{0pt}{\left\{\raisebox{36pt}{}\right.}\hspace{-15pt}&\hspace{0.5pt}\fwbox{115pt}{\fig{-34.76pt}{1}{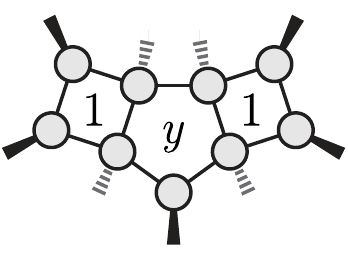}}\hspace{-7.5pt}&\fwbox{105pt}{\fig{-34.76pt}{1}{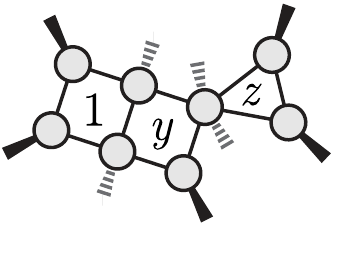}}\hspace{-10pt}&\hspace{-7.5pt}\fwbox{115pt}{\fig{-34.76pt}{1}{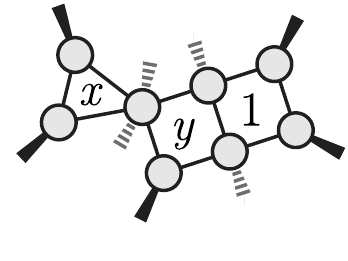}}\fwboxR{0pt}{\left.\raisebox{36pt}{}\right\}}\\[-5.5pt]
\fwboxL{100pt}{\text{{\footnotesize degrees of freedom:}}}&1\hspace{-7.5pt}&1&\hspace{-7.5pt}1
\end{array}}\fwboxR{0pt}{\raisebox{5.35pt}{$(\mathcal{L}_{2})$}}}\nonumber}
\vspace{-5pt}
\eq{\fwbox{0pt}{\fwboxL{430pt}{\begin{array}{@{}c@{}c@{}}\fwbox{100pt}{\fig{-34.76pt}{1}{ladder_int_3}}\hspace{10pt}\fwboxR{0pt}{\left\{\raisebox{29pt}{}\right.}\hspace{-15pt}&\hspace{-2.5pt}\fwbox{125pt}{\fig{-34.76pt}{1}{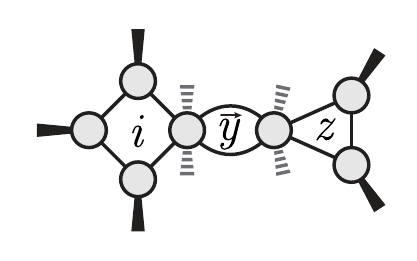}}\fwboxR{0pt}{\left.\raisebox{29pt}{}\right\}}\\[-15.5pt]
\fwboxL{100pt}{\text{{\footnotesize degrees of freedom:}}}&2
\end{array}}\fwboxR{0pt}{\raisebox{0.35pt}{$(\mathcal{L}_{3})$}}}\nonumber}
\vspace{-5pt}
\eq{\fwbox{0pt}{\fwboxL{430pt}{\begin{array}{@{}c@{}c@{}}\fwbox{110pt}{\fig{-34.76pt}{1}{ladder_int_4}}\hspace{10pt}\fwboxR{0pt}{\left\{\raisebox{29pt}{}\right.}\hspace{-15pt}&\hspace{0pt}\fwbox{125pt}{\fig{-34.76pt}{1}{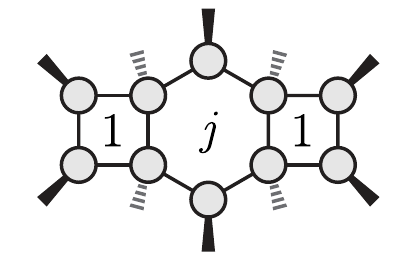}}\fwboxR{0pt}{\left.\raisebox{29pt}{}\right\}}\\[-3.5pt]
\fwboxL{110pt}{\text{{\footnotesize degrees of freedom:}}}&2
\end{array}}\fwboxR{0pt}{\raisebox{6.85pt}{$(\mathcal{L}_{4})$}}}\nonumber}
\vspace{-5pt}
\eq{\fwbox{0pt}{\fwboxL{430pt}{\begin{array}{@{}c@{}c@{}}\fwbox{110pt}{\fig{-34.76pt}{1}{ladder_int_5}}\hspace{10pt}\fwboxR{0pt}{\left\{\raisebox{36pt}{}\right.}\hspace{-15pt}&\hspace{0pt}\fwbox{125pt}{\fig{-34.76pt}{1}{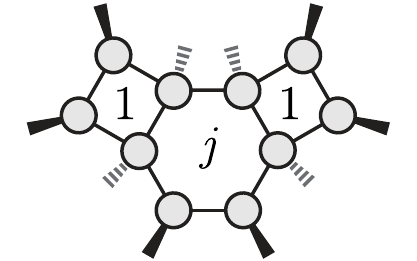}}\fwboxR{0pt}{\left.\raisebox{36pt}{}\right\}}\\[-0.5pt]
\fwboxL{110pt}{\text{{\footnotesize degrees of freedom:}}}&2
\end{array}}\fwboxR{0pt}{\raisebox{7.95pt}{$(\mathcal{L}_{5})$}}}\nonumber}
\vspace{-5pt}
\eq{\fwbox{0pt}{\fwboxL{430pt}{\begin{array}{@{}c@{}c@{}c@{}}\fwbox{110pt}{\fig{-34.76pt}{1}{ladder_int_6}}\hspace{10pt}\fwboxR{0pt}{\left\{\raisebox{36pt}{}\right.}\hspace{-15pt}&\hspace{0.5pt}\fwbox{125pt}{\fig{-34.76pt}{1}{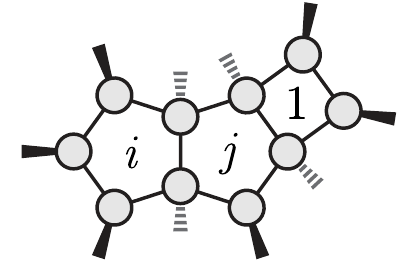}}&\fwbox{120pt}{\fig{-34.76pt}{1}{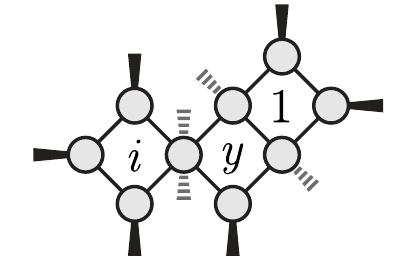}}\fwboxR{0pt}{\left.\raisebox{36pt}{}\right\}}\\[-0.5pt]
\fwboxL{110pt}{\text{{\footnotesize degrees of freedom:}}}&4&2
\end{array}}\fwboxR{0pt}{\raisebox{7.85pt}{$(\mathcal{L}_{6})$}}}\nonumber}
\vspace{-5pt}
\eq{\fwbox{0pt}{\fwboxL{430pt}{\begin{array}{@{}c@{}c@{}}\fwboxL{110pt}{\hspace{-7.5pt}\fig{-34.76pt}{1}{ladder_int_7}}\hspace{10pt}\fwboxR{0pt}{\left\{\raisebox{36pt}{}\right.}\hspace{-15pt}&\hspace{0pt}\fwbox{135pt}{\fig{-34.76pt}{1}{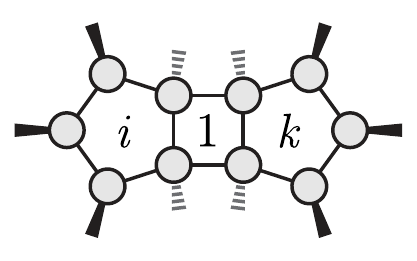}}\fwboxR{0pt}{\left.\raisebox{36pt}{}\right\}}\\[-0.5pt]
\fwboxL{110pt}{\text{{\footnotesize degrees of freedom:}}}&4
\end{array}}\fwboxR{0pt}{\raisebox{8.05pt}{$(\mathcal{L}_{7})$}}}\nonumber}
\vspace{-5pt}
\eq{\fwbox{0pt}{\fwboxL{430pt}{\begin{array}{@{}c@{}c@{}}\fwboxL{120pt}{\fig{-34.76pt}{1}{ladder_int_8}}\hspace{10pt}\fwboxR{0pt}{\left\{\raisebox{36pt}{}\right.}\hspace{-15pt}&\hspace{0pt}\fwbox{135pt}{\hspace{-5pt}\fig{-34.76pt}{1}{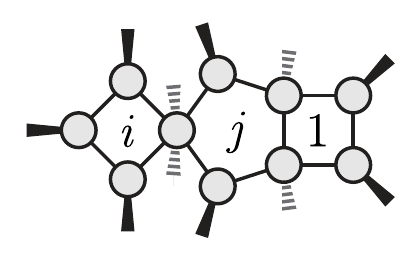}}\fwboxR{0pt}{\left.\raisebox{36pt}{}\right\}}\\[-5.5pt]
\fwboxL{120pt}{\text{{\footnotesize degrees of freedom:}}}&4
\end{array}}\fwboxR{0pt}{\raisebox{5.85pt}{$(\mathcal{L}_{8})$}}}\nonumber}
\vspace{-5pt}
\eq{\fwbox{0pt}{\fwboxL{430pt}{\begin{array}{@{}c@{}c@{}}\fwboxL{120pt}{\fig{-34.76pt}{1}{ladder_int_9}}\hspace{10pt}\fwboxR{0pt}{\left\{\raisebox{36pt}{}\right.}\hspace{-15pt}&\hspace{0pt}\fwbox{125pt}{\fig{-34.76pt}{1}{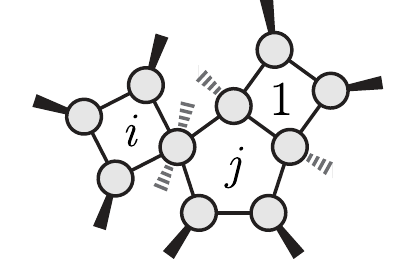}}\fwboxR{0pt}{\left.\raisebox{36pt}{}\right\}}\\[-4.5pt]
\fwboxL{120pt}{\text{{\footnotesize degrees of freedom:}}}&4
\end{array}}\fwboxR{0pt}{\raisebox{6.05pt}{$(\mathcal{L}_{9})$}}}\nonumber}
\vspace{-5pt}
\eq{\fwbox{0pt}{\fwboxL{430pt}{\begin{array}{@{}c@{}c@{}c@{}}\fwbox{120pt}{\fig{-34.76pt}{1}{ladder_int_10}}\hspace{10pt}\fwboxR{0pt}{\left\{\raisebox{36pt}{}\right.}\hspace{-15pt}&\hspace{0.5pt}\fwbox{130pt}{\hspace{-10pt}\fig{-34.76pt}{1}{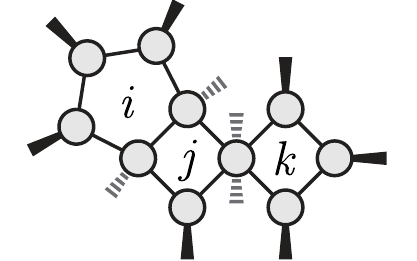}}&\fwbox{65pt}{\hspace{10pt}\fig{-34.76pt}{1}{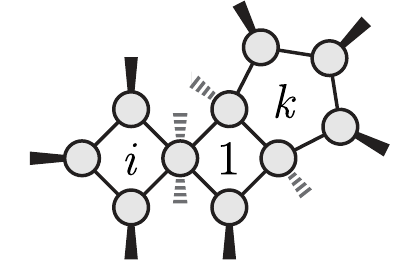}\fwboxR{0pt}{\left.\raisebox{36pt}{}\right\}}}\\[-0.5pt]
\fwboxL{120pt}{\text{{\footnotesize degrees of freedom:}}}&8&4
\end{array}}\fwboxR{0pt}{\raisebox{7.55pt}{$(\mathcal{L}_{10})$}}}\nonumber}
\vspace{-5pt}
\eq{\fwbox{0pt}{\fwboxL{430pt}{\begin{array}{@{}c@{}c@{}}\fwboxL{120pt}{\fig{-34.76pt}{1}{ladder_int_11}}\hspace{10pt}\fwboxR{0pt}{\left\{\raisebox{36pt}{}\right.}\hspace{-15pt}&\hspace{0pt}\fwbox{135pt}{\hspace{-0pt}\fig{-34.76pt}{1}{ladder_cut_11a}}\fwboxR{0pt}{\left.\raisebox{36pt}{}\right\}}\\[-5.5pt]
\fwboxL{120pt}{\text{{\footnotesize degrees of freedom:}}}&8
\end{array}}\fwboxR{0pt}{\raisebox{5.35pt}{$(\mathcal{L}_{11})$}}}\nonumber}
\vspace{-5pt}
\eq{\fwbox{0pt}{\fwboxL{430pt}{\begin{array}{@{}c@{}c@{}}\fwboxL{120pt}{\fig{-34.76pt}{1}{ladder_int_12}}\hspace{10pt}\fwboxR{0pt}{\left\{\raisebox{36pt}{}\right.}\hspace{-15pt}&\hspace{0pt}\fwbox{115pt}{\fig{-34.76pt}{1}{ladder_cut_12a}}\fwboxR{0pt}{\left.\raisebox{36pt}{}\right\}}\\[-5.5pt]
\fwboxL{120pt}{\text{{\footnotesize degrees of freedom:}}}&8
\end{array}}\fwboxR{0pt}{\raisebox{5.35pt}{$(\mathcal{L}_{12})$}}}\nonumber}

~\vspace{40pt}
\providecommand{\href}[2]{#2}\begingroup\raggedright\endgroup

\end{document}